%% file: main.tex
\pdfoutput=1
\documentclass[sigconf]{acmart}
\AtBeginDocument{%
  }

\copyrightyear{2026}
\acmYear{2026}
\setcopyright{cc}
\setcctype{by}
\acmConference[CHI '26]{Proceedings of the 2026 CHI Conference on Human Factors in Computing Systems}{April 13--17, 2026}{Barcelona, Spain}
\acmBooktitle{Proceedings of the 2026 CHI Conference on Human Factors in Computing Systems (CHI '26), April 13--17, 2026, Barcelona, Spain}
\acmDOI{10.1145/3772318.3791403}
\acmISBN{979-8-4007-2278-3/2026/04}

\usepackage{xcolor}   
\usepackage[normalem]{ulem}   
\usepackage{booktabs}  
\usepackage{multirow}  
\usepackage{graphicx}
\usepackage{url}
\usepackage{enumitem}
\usepackage{microtype}
\usepackage{multirow}
\usepackage[utf8]{inputenc}
\usepackage{balance}
\usepackage{bm}
\usepackage{balance}
\newcommand{\rv}[1]{\textcolor[rgb]{0, 0, 0}{#1}}  

\begin{document}





\title{Beyond Input--Output: Rethinking Creativity through Design-by-Analogy in Human--AI Collaboration}

\settopmatter{authorsperrow=4}

\author{\texorpdfstring{\textbf{Xuechen Li} \vspace{5pt}}{Xuechen Li}}
\affiliation{%
  \institution{College of Design and Innovation,}
  \institution{Tongji University}
  \city{Shanghai}
  \country{China}}
\email{primav1022@tongji.edu.cn}

\author{\texorpdfstring{\textbf{Shuai Zhang} \vspace{5pt}}{Shuai Zhang}}
\affiliation{%
  \institution{College of Design and Innovation,}
  \institution{Tongji University}
  \city{Shanghai}
  \country{China}}
\email{zhangshuai2@tongji.edu.cn}

\author{\texorpdfstring{\textbf{Nan Cao} \vspace{5pt}}{Nan Cao}}
\affiliation{%
  \institution{College of Design and Innovation,}
  \institution{Tongji University}
  \city{Shanghai}
  \country{China}}
\email{caonan@tongji.edu.cn}


\author{Qing Chen}
\authornote{Qing Chen is the corresponding author.}
\affiliation{%
  \institution{College of Design and Innovation,}
  \institution{Tongji University}
  \city{Shanghai}
  \country{China}}
\email{qingchen@tongji.edu.cn}



\newcommand{\cq}[1]{{\color{blue}#1}}
\renewcommand{\shortauthors}{Li et al.}

\begin{abstract}

While the proliferation of foundation models has significantly boosted individual productivity, it also introduces a potential challenge: the homogenization of creative content~\cite{doshi2024generative}. In response, we revisit Design-by-Analogy (DbA), a cognitively grounded approach that fosters novel solutions by mapping inspiration across domains. However, prevailing perspectives often restrict DbA to early ideation or specific data modalities, while reducing AI-driven design to simplified input–output pipelines. Such conceptual limitations inadvertently foster widespread design fixation. To address this, we expand the understanding of DbA by embedding it into the entire creative process, thereby demonstrating its capacity to mitigate such fixation. Through a systematic review of 85 studies, we identify six forms of representation and classify techniques across seven stages of the creative process. We further discuss three major application domains: creative industries, intelligent manufacturing, and education and services, demonstrating DbA’s practical relevance. Building on this synthesis, we frame DbA as a mediating technology for human-AI collaboration and outline the potential opportunities and inherent risks for advancing creativity support in HCI and design research.

\end{abstract}

\begin{teaserfigure}
\includegraphics[width=\textwidth]{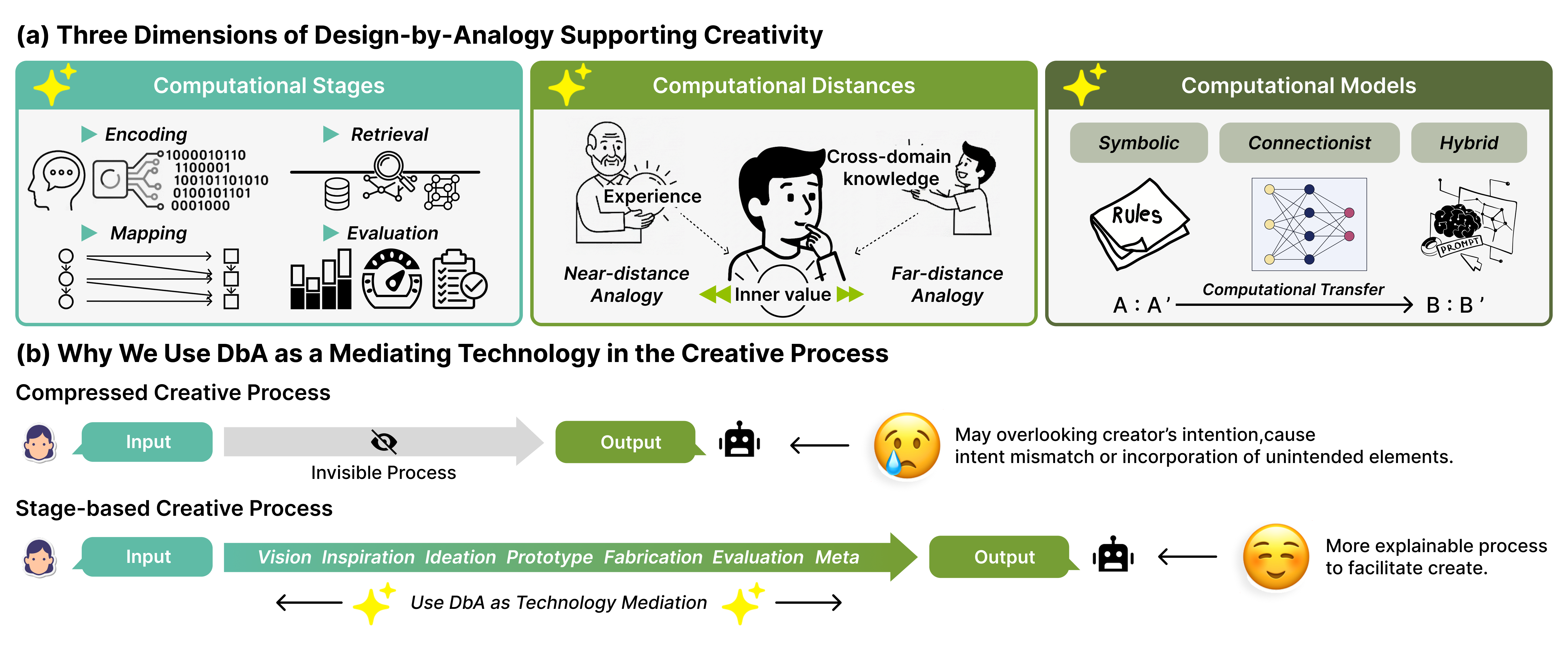}
\caption{This figure (a) illustrates the three dimensions through which the Design-by-Analogy (DbA) method supports creativity. It also explains (b) why DbA could be served as a technology mediation~\cite{smits2019values} in the creative process, forming the foundation of this paper.}
\Description{}
\label{fig:Teaser}
\end{teaserfigure}

\ccsdesc[500]{Human-centered computing~HCI theory, concepts and models}
\ccsdesc[500]{Human-centered computing~Interaction paradigms}
\ccsdesc[500]{Human-centered computing~Collaborative interaction}


\keywords{Design-by-Analogy, Creativity Support Tool, Design Method, Systematic Review}


\maketitle
\input{sections/1_introduction}

\input{sections/2_related_work}
\input{sections/3_methodology}
\input{sections/4_Representation}
\input{sections/5_Creative_Process}

\input{sections/6_Application}

\input{sections/7_Findings}
\input{sections/8_Discussion}
\flushbottom 
\input{sections/9_Conclusion}
\input{sections/10_Acknowledgment}

\bibliographystyle{ACM-Reference-Format}
\bibliography{ref}

\end{document}

%% file: sections/1_introduction.tex
\section{Introduction}

Recent advances in foundational models have transformed creative work by rapidly generating text, images, and designs at scale. Yet their efficiency comes with a cost: by standardizing creative expression, they risk diminishing the collective diversity of ideas and solutions~\cite{doshi2024generative}. Addressing this issue requires not only technical improvements but also a deeper understanding of how human creativity operates and how it can be meaningfully supported in practice. \rv{Looking back on the past, the study of creativity has evolved from philosophical origins to a multidisciplinary science, significantly advanced by 20th-century psychology through concepts like divergent thinking~\cite{mccrae1987creativity}, synectics~\cite{gordon1961synectics}. Subsequent cognitive approaches systematically modeled the creative process, introducing the notion of ``analogy'' and ``metaphor''~\cite{holyoak1996mental}. With the rise of computational methods~\cite{kenett2023semantic}, and more recently generative AI, creativity support systems have become a tangible prospect.}
However, translating these insights into pragmatic design tools remains a challenge for real-world design and human-computer interaction (HCI). 

Among these mechanisms, Design-by-Analogy (DbA) provides a powerful bridge between cognition and practice. DbA generates novel solutions by drawing inspiration from a source domain and mapping it to a target domain~\cite{jiang2022data}. DbA has roots in Gentner's Structure-Mapping Theory~\cite{gentner1983structure} and has been applied in various contexts, such as bio-inspired design~\cite{linsey2013overcoming}, mechanical design~\cite{jiang2022data}, and patent-based design~\cite{verhaegen2011identifying}. Real-world examples, such as wind turbine blades inspired by humpback whale tubercles~\cite{fu2014bio}, demonstrate its potential to expand creative horizons.


Despite these advances, existing DbA reviews have predominantly focused on theoretical frameworks~\cite{linsey2008increasing}; data-driven approaches with limited sources~\cite{jiang2022data}; applications of single computational techniques, \rv{including case-based reasoning approaches~\cite{aamodt1994case} and deep generative machine learning models~\cite{ regenwetter2022deep}; as well as interaction-centered methods such as the role of timing and analogical similarity~\cite{tseng2008role} or different forms of representation~\cite{ marshall2016analogy} in the process of idea generation}. While each strand is valuable, this specialized focus fosters implicit assumptions that constrain the perceived scope of DbA, confining it to early-stage ideation or specific data modalities. 
Concurrently, a reductionist trend in industry, especially in AI-driven design, oversimplifies creativity into input-output models~\cite{chakrabarty2024art, wadinambiarachchi2024effects}, which can lead to design fixation and the compression of the designer's intentions. \rv{We hypothesised that embedding DbA systems within the creative process reframes the role of AI from solution-providers to cognitive guides. By intervening at critical junctures, such systems utilize AI's analogical reasoning to foster user creativity and, at the same time, moderate the prescriptive nature of generated content.} Together, these issues highlight the need for a holistic perspective on DbA that bridges cognitive science, design research, and HCI. 

To address these gaps, we conducted a systematic review using a PRISMA selection process~\cite{page2021prisma} of research on DbA and related fields. From an initial corpus of 1615 publications, we identified and analyzed 85 articles that met our inclusion criteria. This corpus covers technologies, applications, and theoretical contributions across domains. 

Our analysis reveals that DbA manifests in six distinct representation forms:
\textit{\textbf{Semantics and Text}}, \textit{\textbf{Visual and Appearance}}, \textit{\textbf{Material and Structure}}, \textit{ \textbf{Function and Attribute}}, \textit{\textbf{Interaction, Workflows, and Multi-sensory experience}}, and \textit{\textbf{Unconventional Contexts}}. 
These forms span all seven stages in the four-phase creative process:
\textbf{Problem Definition} (vision, inspiration), \textbf{Product Ideation} (ideation, prototype), \textbf{Product Implementation} (fabrication), and \textbf{Product Evaluation} (evaluation, meta). In addition, we highlight DbA's applicability across three major domains: \textbf{Creative Industries}, \textbf{Intelligent Manufacturing}, \textbf{Education and Service Industries}.
Through our systematic review and discussion of application domains, this study aims to answer three research questions: 
(1) How can DbA be articulated as a mediating technique that activates human cognitive potential in practice?
(2) What is the current landscape of AI-driven DbA technologies, and how do they support human–AI collaboration across different creative stages?
\rv{(3) What references and opportunities can be concluded from DbA in HCI and design research?}


The contributions of this work are threefold: 


(1) \textbf{Systematic Review and Corpus Construction.} We curate and analyze a comprehensive corpus of DbA research to provide a foundation for HCI and design researchers exploring related studies.\rv{ (RQ1-3)}

(2) \textbf{\rv{Representation Taxonomy and Technique Classification.}} We identify six representational forms and map them to creative processes and domains, offering a conceptual lens for designing interactive DbA-support tools. \rv{(RQ1-2)}

(3) \textbf{Guidelines and Future Directions.} We propose guidelines and research opportunities for AI-supported DbA, highlighting implications for HCI research on creativity support, human-AI collaboration, and design practice.\rv{(RQ3)}

%% file: sections/2_related_work.tex
\section{Background}

\subsection{Design-by-Analogy in Creativity Support}


\rv{The study of creativity has evolved significantly from its ancient philosophical origins~\cite{simonton1999origins}, where it was regarded as a divine gift, to its current status as a multidisciplinary scientific field. During the 20th century, psychology was instrumental in driving this transformation. The psychometric school, exemplified by prominent figures such as J. P. Guilford~\cite{guilford1967creativity} and W.J.J. Gordon~\cite{gordon1961synectics}, marked a pivotal turning point through the operationalisation of creativity as a measurable variable. This development enabled empirical studies by introducing concepts such as divergent thinking~\cite{guilford1967creativity} and Synectics~\cite{gordon1961synectics}. Subsequently, cognitive psychology undertook a systematic analysis of the creative process~\cite{gentner1983structure}, decomposing it into distinct stages: problem identification, information encoding, knowledge retrieval, and conceptual reorganization. Furthermore, Gestalt psychology introduced the concept of ``insight''\cite{kohler1967gestalt}. In neuroscience, the scientific study of the human brain has identified bilateral frontal activation as a neural correlate of creativity~\cite{carlsson2000neurobiology}. In contrast, ecological and systems approaches have theorised that creativity is an emergent property resulting from the interaction between individuals, domains, and fields~\cite{finke1996imagery, simonton2012fields}. By the early 21st century, computational methods for simulating creative processes began to emerge~\cite{french2002computational, gentner2011computational}. Recent developments, particularly the rise of generative AI, have prompted deeper reflection on human creativity and empathy~\cite{becue2024ai}, making AI-driven creative support a foreseeable prospect. Design-by-Analogy (DbA) stands as a derivative of Synectics~\cite{gordon1961synectics}, deeply rooted in the history of creativity support methods. Benefiting from the emergent capabilities of generative AI, DbA has found new soil for its innovative development.}


DbA facilitates the generation of novel solutions by transferring cross-domain insights from a source to a target domain~\cite{jiang2022data}. The conceptual framework of DbA originated with Gentner's Structure-Mapping theory in 1983~\cite{gentner1983structure}, which laid the foundation for decades of research on computational analogy and evaluation metrics~\cite{thagard1990analog, hofstadter1984copycat}. In 2008, Linsey et al. introduced the WordTree method to further advance DbA~\cite{linsey2008increasing}, leading to diverse applications in mechanical design~\cite{fu2015design}, bio-inspired design~\cite{fu2014bio}, and conceptual design~\cite{moreno2016overcoming}. This evolution established DbA as a practical design methodology rather than a purely cognitive theory.
Unlike analogy in pure cognitive science, DbA situates its reasoning within the pragmatic context of design, characterized by Simon as the process of \textit{``transforming existing conditions into preferred ones''}~\cite{simon2019sciences}. In this way, DbA operationalizes analogy as a goal-driven process that bridges abstract reasoning with tangible outcomes, systematically activating designers' memory, inspiration, and intrinsic motivation~\cite{moreno2015step, schecter2025role}.


The efficacy of DbA in supporting creativity stems from its unique operational mechanism, as illustrated by the three dimensions in Fig.\ref{fig:Teaser}. First, it decomposes the analogy-making component of the creative process into four computational stages: representation, retrieval, mapping, and evaluation~\cite{jiang2022data, fu2014bio, fu2015design}. Second, DbA enables the transfer of knowledge across varying distances, ranging from near-domain experience to far cross-domain inspiration, all while remaining guided by design goals~\cite{tseng2008role} and human values~\cite{schecter2025role, chan2011benefits}. Finally, its structured logic makes creativity computationally tractable (i.e., ``\textbf{A : A' as B : B'}''). Various approaches, including symbolic, connectionist, and hybrid AI models, have been employed to implement mapping and representation~\cite{thagard1990analog, hummel1997distributed, hofstadter1984copycat}. Together, these properties allow DbA to mitigate design fixation and guide designers beyond conventional solutions without compromising human agency~\cite{moreno2015step, goucher2019neuroimaging}.


Compared to other computational creativity support methods, DbA offers several distinct advantages. It is significantly more effective than unstructured brainstorming in converting novel ideas into practical solutions~\cite{linsey2010study}, and its structured framework reduces cognitive load relative to pure analogical reasoning~\cite{richland2013reducing}. In contrast with the highly automated retrieve-reuse-revise-retain paradigm of Case-Based Reasoning (CBR), DbA preserves greater designer agency and fosters intrinsic motivation, ensuring that computational support enhances rather than replaces human creativity~\cite{wills1994towards}.


\subsection{Related Surveys}
Design-by-Analogy (DbA) has demonstrated broad applicability, extending from bio-inspired innovations, such as radar systems based on bat echolocation~\cite{helms2009biologically} and wind turbine blades mimicking whale fins~\cite{linsey2008increasing}, to everyday product design~\cite{linsey2012design, barnett2002and}. Beyond engineering, its underlying principles are also investigated as forms of metacognition~\cite{ball2019advancing} across education, philosophy, and mathematics~\cite{holyoak1996mental}.

Existing surveys of DbA research generally coalesce around three interconnected themes. The first focuses on \textbf{knowledge representation and retrieval}, which serves as the foundational bedrock of DbA. Key contributions in this area include data-driven DbA surveys~\cite{jiang2022data}, investigations into data representations and modalities~\cite{linsey2008modality, linsey2010study}, taxonomies of analogical design~\cite{fu2014bio}, and applications of biological inspiration to practical design challenges~\cite{o2015toward}.

The second theme centers on \textbf{computational models and synthesis} designed to assist or automate the creative process. Representative scholarship encompasses computer-based design synthesis~\cite{chakrabarti2011computer}, Semantic Theory of Inventive Problem Solving (TRIZ) Techniques~\cite{ghane2024semantic}, deep generative models for engineering design~\cite{regenwetter2022deep}, and knowledge recommendation for 4D printing~\cite{dimassi2023knowledge}. Further developments include analogical retrieval tools based on the structure-mapping function model~\cite{liu2023smfm} and case-based reasoning frameworks~\cite{aamodt1994case}. More recent studies have also begun to map various AI methodologies to specific stages of the design process~\cite{khanolkar2023mapping}.

The third theme investigates \textbf{human-computer interaction and cognition}. Research in this area examines how analogical design affects creative processes and fixation~\cite{atilola2015representing, marshall2016analogy}, how cognitive science informs computer-aided systems~\cite{goel2012cognitive}, and metrics for evaluating analogical ideas~\cite{verhaegen2013refinements, lu2023differences}. These studies highlight the importance of timing, representation, and context in supporting human creativity.



Despite this extensive body of research, significant gaps remain. Existing reviews often concentrate on specific disciplines, data types, or isolated techniques~\cite{jiang2022data, khanolkar2023mapping, ghane2024semantic}, thereby lacking a unified framework for understanding DbA across diverse representations, creative processes, and application domains. To address these limitations, this paper presents a systematic review centered on the intersection of computational creativity and human creative behavior. By analyzing the representations, techniques, and applications of DbA, we aim to demonstrate how this methodology can augment creativity and reinforce human values in the age of AI.



%% file: sections/3_methodology.tex
\section{Methodology} 
\begin{figure*}[t]
    \centering
    \includegraphics[width=1\linewidth]{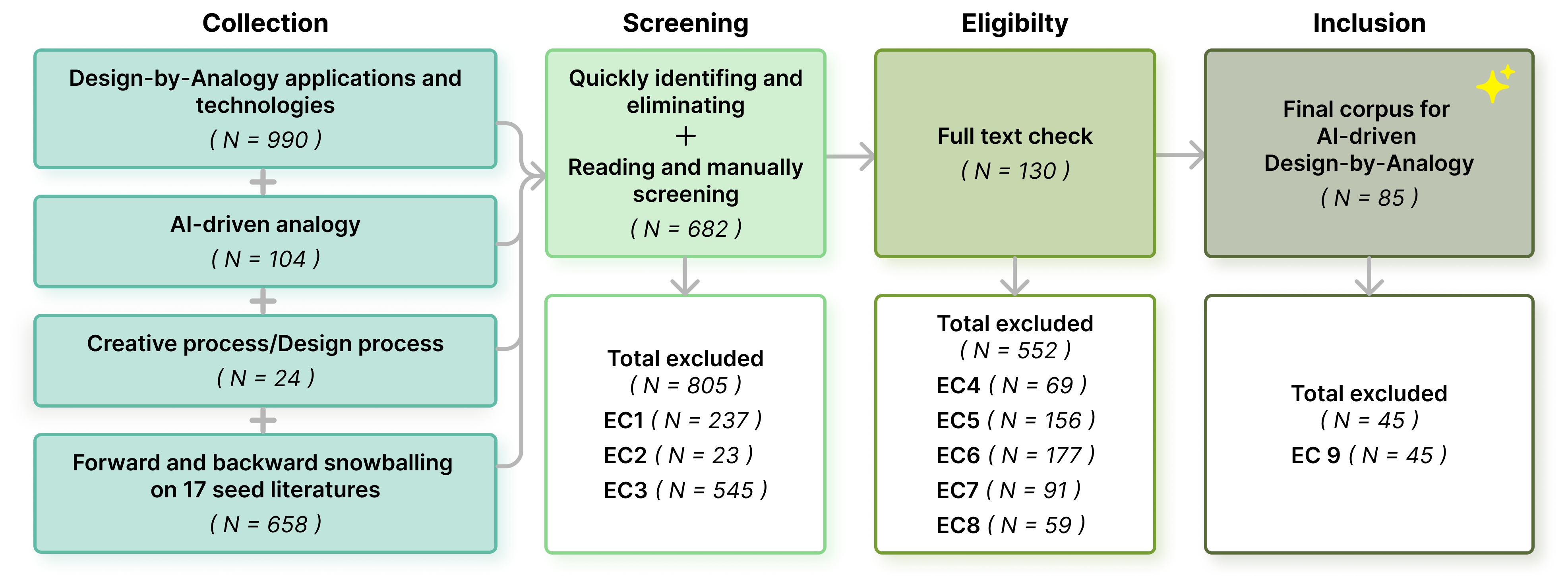}
    \caption{Overview of the PRISMA selection process used to construct the final corpus.}
    \Description{}
    \label{fig:prisma-selection}
\end{figure*}

To systematically review AI-driven DbA, we adopted the PRISMA guidelines ~\cite{page2021prisma}, ensuring transparency and rigor throughout the literature selection process (Fig. \ref{fig:prisma-selection}).


\subsection{Data Collection}

In the \textbf{identification stage}, we initiated the process with both programmatic and manual searches across core disciplines, including HCI, computer science, design studies, cognitive science, and mechanical design. \rv{We conducted a comprehensive search across major databases, ACM Digital Library, IEEE Xplore, ASME Digital Collection, and ScienceDirect, to cover all relevant academic contributions.} Our initial collection included 990 records on DbA applications and technologies, 104 on AI-driven analogy, and 24 on creative process taxonomies. From these, 17 \rv{initial benchmark papers} with high citation rates and relevance were identified. We then performed forward ($N = 30$) and backward ($N = 30$) snowballing, yielding 658 additional works. In total, 1,615 records were retrieved using keywords such as ``\textit{Design-by-analogy}'', ``\textit{Analogy Reasoning}'', ``\textit{Analogical Design}'', ``\textit{Case-based Reasoning}'', and ``\textit{Analogy Making}''. After removing duplicates, 1,378 unique items remained.


\subsection{PRISMA Selection}

In the \textbf{screening stage}, we applied a programmatic approach to efficiently identify and eliminate literature clearly unrelated to the topic based on titles and abstracts. \rv{We established exclusion criteria (EC) to filter the dataset: excluding literature with duplicate titles} (\textbf{EC1}, $N = 237$), non-peer-reviewed works (\textbf{EC2}, $N = 23$), \rv{and studies where DbA was mentioned solely as an illustrative example or motivation rather than the core subject} (\textbf{EC3}, $N = 545$). This process resulted in 682 candidate papers.

\begin{figure*}[t]
    \centering
    \includegraphics[width=1\linewidth]{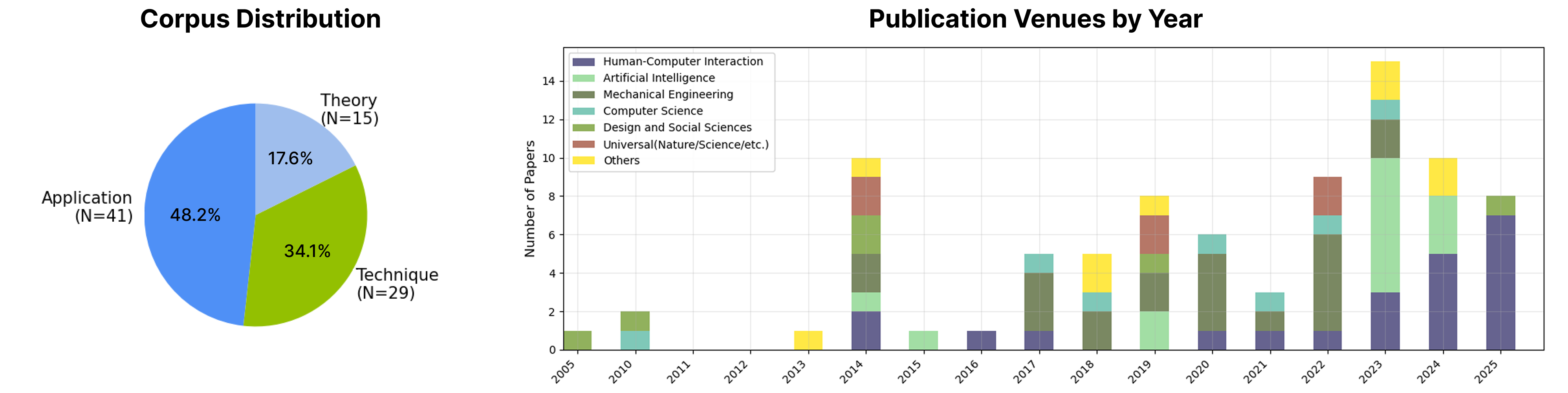}
    \caption{Statistical overview of the 85 paper corpus. The pie chart on the left illustrates the distribution by research focus, with application-oriented studies (48.2\%) forming the largest category. The stacked bar chart on the right displays the temporal distribution of publications from 2005 to 2025, with colors representing the breakdown by venue categories, such as Human-Computer Interaction and Artificial Intelligence. Detailed corpus information can be found in the \textit{Appendix.}}
    \Description{}
    \label{fig:statistics}
\end{figure*}

In the \textbf{eligibility assessment stage}, two researchers conducted multiple rounds of discussion to establish explicit exclusion criteria for full-text screening. We removed studies that: lacked integration with design processes, where analogy served merely as an isolated algorithmic component (\textbf{EC4}, $N = 69$); did not directly address creative work (e.g., pure mathematics or medical diagnosis) (\textbf{EC5}, $N = 156$); utilized analogy without specific mapping details (\textbf{EC6}, $N = 177$); validated DbA mechanisms empirically without proposing concrete methods or techniques (\textbf{EC7}, $N = 91$); or relied solely on non-AI computational methods (\textbf{EC8}, $N = 59$). This screening left 130 papers.

In the \textbf{inclusion stage}, thematic analysis~\cite{terry2017thematic} was applied to ensure coverage across theory, technology, and application. Two authors independently analyzed the screened literature against our taxonomy system, engaging in iterative discussions to refine the selection. \rv{After communication, the two authors collectively excluded the articles on which they had inconsistent opinions regarding the contributions to the DbA field (\textbf{EC9}, $N = 45$).} Consequently, 85 publications were retained for further analysis.


\textbf{Limitations.} The initial search employed keyword retrieval based on specific definitions and criteria, which inherently entails a degree of subjectivity. Furthermore, the 85 retained references primarily encompass DbA methodologies focused on domain-specific scenarios. Applications addressing emerging scenarios and interdisciplinary frontiers practice, such as the circular economy, remain areas requiring further exploration.

\subsection{Validation and Corpus Overview}

To validate corpus reliability, two HCI researchers independently annotated the dataset using a 17-dimensional coding guide across three major categories. This guide provided definitions, judgment criteria, and boundary conditions for each dimension, with any discrepancies resolved through discussion. We quantitatively assessed inter-rater reliability (IRR) using Cohen's Kappa coefficient. The results demonstrated substantial agreement across all categories: ``Analogy Process'' (including encoding, retrieval, mapping, and evaluation) achieved a composite Kappa ($\kappa = 0.6312$), ``Creative Process'' achieved a composite ($\kappa = 0.7967$), and ``Analogy Type'' yielded a composite ($\kappa = 0.7593$). The overall average Kappa value was $\kappa = 0.7289$.

The final corpus of 85 publications spans a diverse collection of AI-driven DbA research (Fig.(~\ref{fig:statistics})). Regarding research focus, works were application-oriented (48.2\%, $N = 41$), technology-focused (35.3\%, $N = 29$), or theoretical (17.6\%, $N = 15$). In terms of domain, studies addressed intelligent manufacturing (41.2\%, $N = 35$), creative industries (32.9\%, $N = 28$), and education and services (27.1\%, $N = 23$). Based on automation levels~\cite{parasuraman2000model}, systems were classified as: \textit{Assist} (35.3\%, $N = 30$), \rv{where AI and computational tools reduce cognitive load through retrieval or mapping (LoA 2--4)}; \textit{Augment} (38.8\%, $N = 33$), \rv{where humans and AI jointly expand creative exploration (LoA 5)}; and \textit{Automate} (27.1\%, $N = 23$), \rv{where systems generate and iterate content autonomously with human guidance (LoA 6--7)}. The corpus includes work from top-tier HCI conferences (e.g., CHI, UIST, TEI, DIS), leading engineering journals (e.g., JMD, AI EDAM), and high-impact interdisciplinary journals (e.g., Nature, Science), reflecting a broad research landscape.



%% file: sections/4_Representation.tex
\section{Representations in Design-by-Analogy} 

\begin{figure*}[t] 
    \centering
    \includegraphics[width=1\linewidth]{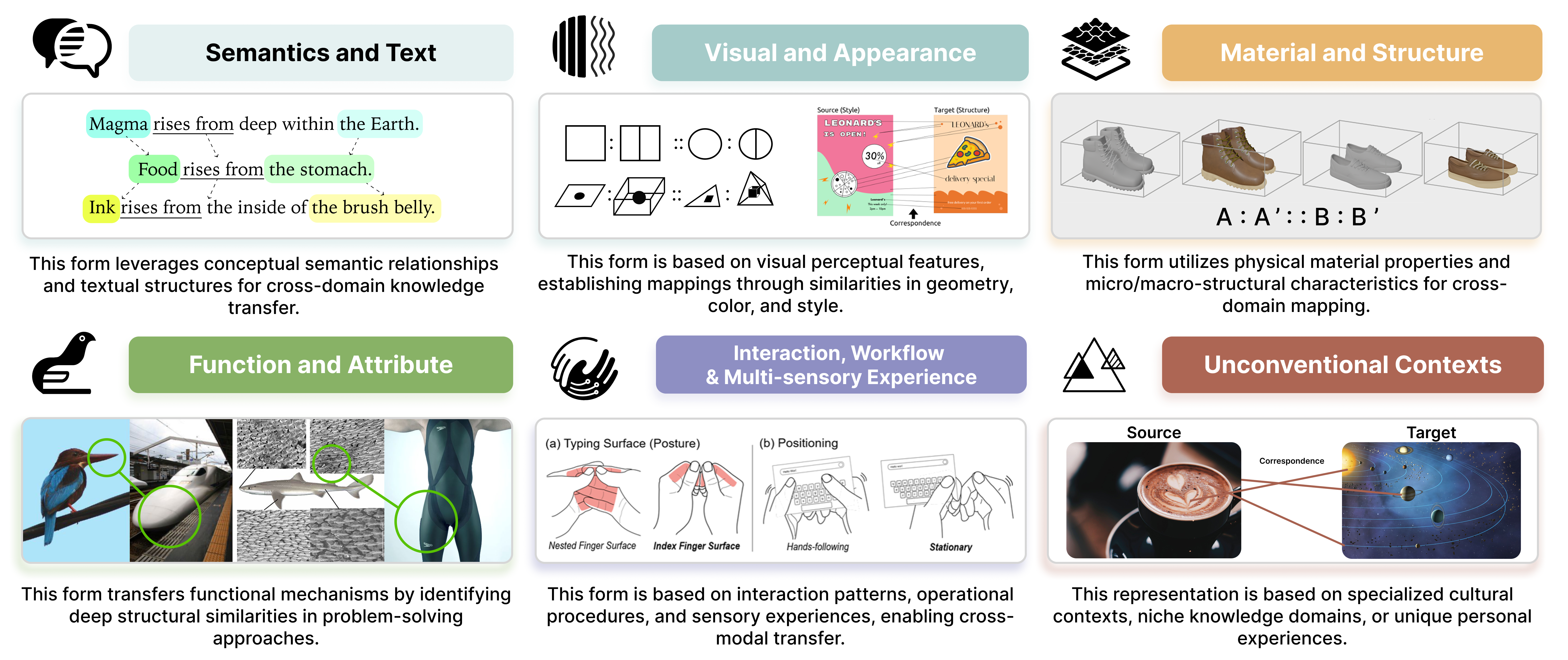}
    \caption{A taxonomy of six representations for Design-by-Analogy: (1) Semantics \& Text~\cite{jiayang2023storyanalogy}, (2) Visual \& Appearance~\cite{warner2023interactive}, (3) Material \& Structure~\cite{fischer2024nerf}, (4) Function \& Attribute~\cite{fu2014bio}, (5) Interaction \& Experience~\cite{kim2023star}, and (6) Unconventional Contexts~\cite{gentner1983structure, linsey2008modality}.}
    \Description{}
    \label{fig:Representation}
\end{figure*}

Drawing from prior work in design research~\cite{hertzmann2023image, linsey2008modality} and analogy theory~\cite{gentner1983structure}, we identify six forms of representation in DbA, as illustrated in Fig.~\ref{fig:Representation}. Representation is foundational to DbA; the manner in which knowledge is encoded in the source domain directly shapes its transferability and creative potential. \rv{Within our collected corpus, the distribution of these representations is as follows: \textit{Function \& Attribute} (54.1\%, $N=46$), \textit{Semantics \& Text} (40.0\%, $N=34$), \textit{Interaction \& Experience} (31.8\%, $N=27$), \textit{Material \& Structure} (28.2\%, $N=24$), \textit{Visual \& Appearance} (21.2\%, $N=18$), and \textit{Unconventional Contexts} (17.6\%, $N=15$).}


\underline{Semantics and Text.} Text-based representations leverage semantic structures and linguistic relations to facilitate analogy. They enable the similarity-based generation of empathic stories or scientific narratives~\cite{shao2025unlock, Ju2025toward}, crowd-sourced idea combination~\cite{srinivasan2024improving, yu2014distributed}, and cross-disciplinary mappings through knowledge retrieval~\cite{zheng2024disciplink}. Furthermore, they support the development of text-centric knowledge repositories for domains such as engineering~\cite{you2018design}, service design~\cite{moreno2014analogies}, and military systems~\cite{yucelmics2018procedure}.

\underline{Visual and Appearance.} Visual analogies rely on perceptual similarities in geometry, color, and style. Key applications encompass vector graphic style transfer~\cite{warner2023interactive}, sketch migration~\cite{lin2025inkspire}, and 3D model analogies~\cite{fischer2024nerf}. Beyond direct transfer, other approaches mine design concepts based on visual resemblance~\cite{yu2016distributed, edwards2024advise}, interpret metaphors via visual context mapping~\cite{akula2023metaclue}, or detect overlooked structural similarities~\cite{peyre2019detecting}.

\underline{Material and Structure.} This category utilizes physical properties and structural characteristics for cross-domain mapping, distinguishing between two primary subgroups. The first facilitates transfer through similarities in physical properties, such as material finish, strength, and thermal conductivity. Examples include material migration in product design~\cite{fischer2024nerf, khosravani2022intelligent}, mapping mechanical fatigue to monitoring visualizations~\cite{thomas2013extending}, and developing flexible electronics inspired by biological tissue~\cite{hong2024fishbone}. The second subgroup derives from microstructural analogies based on geometric topology, encompassing fractal analogies for stretchable devices~\cite{fan2014fractal} and 2D-to-3D structural transformations~\cite{jin2023deep}.

\underline{Function and Attribute.} Functional analogies identify structural or relational similarities to aid in problem-solving. First, the core functional consistency aligns, mapping the functions that address similar problems, such as analogical database retrieval~\cite{murphy2014function}, software feature recommendation~\cite{zhang2017systematic}, and solution mapping from knowledge bases~\cite{khosravani2022intelligent}. Second, the transfer of principles from domain to domain, applying knowledge from distinct fields, illustrated by transferring biological functions to mechanics~\cite{kang2025biospark}, optimizing energy use based on human behavior~\cite{gonzalez2018energy}, or scaling innovation through AI and crowdsourcing~\cite{kittur2019scaling}. Finally, relational equivalence, maps structural relationships; for example, employing proportional text or imagery to communicate complex data~\cite{chen2024beyond}, explaining quantum concepts through AR~\cite{karunathilaka2025intuit}, or mapping craft tutorials based on material practices~\cite{emerson2024anther}.

\underline{Interaction, Workflows, and Multi-sensory Experience.} This representation focuses on operational flows and sensory patterns. First, interaction similarity maps graphic software interactions to text editing~\cite{masson2025textoshop} or transfers interaction paradigms between websites~\cite{chen2021umitation}. Second, workflow similarity aligns manufacturing processes with ontologies~\cite{hagedorn2018knowledge} or maps robotic workflows to chemical synthesis~\cite{coley2019robotic}. Finally, sensory experience mapping translates tactile, auditory, or kinesthetic sensations across contexts~\cite{kim2023star,wang2024reelframer,la2020designing}.


 
\underline{Unconventional Contexts.} This form draws upon specialized cultural contexts, niche knowledge domains, or unique personal experiences. Prominent examples include embedding Tai Chi culture into drone technologies~\cite{la2020designing}, adapting perfume manufacturing techniques to wine production through local knowledge systems~\cite{andriani2025perfume}, compiling domain-specific terminology into personalized lexicons for cross-disciplinary collaboration~\cite{cao2025medai}, and leveraging the diverse expertise of team members to stimulate collaborative design~\cite{christensen2016creative}.

%% file: sections/5_Creative_Process.tex
\section{Design-by-Analogy Techniques in the Creative Process}
\begin{figure*}[t]
    \centering
    \includegraphics[width=1\linewidth]{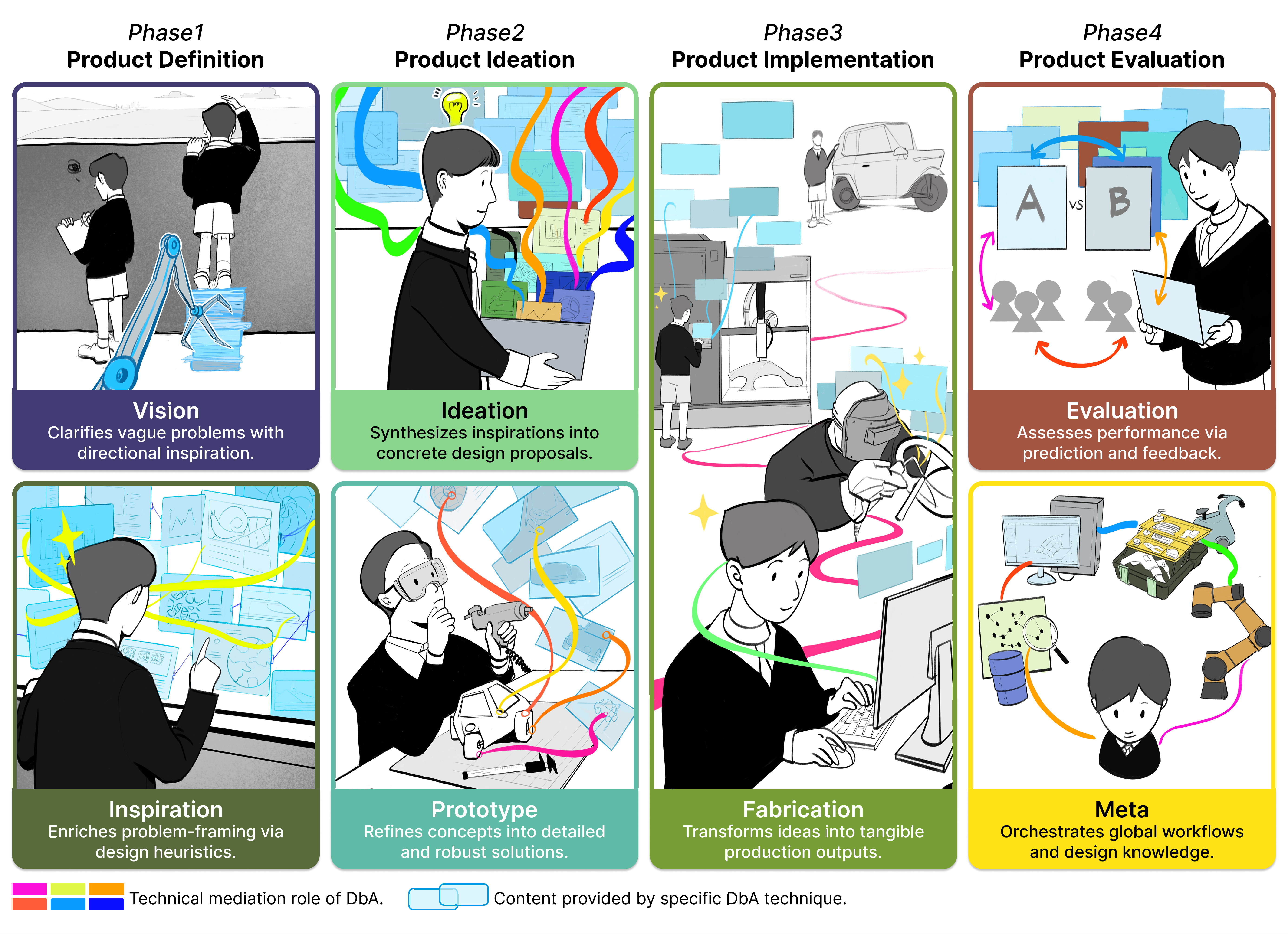}
    \caption{From the opening image (Phase 1: Vision) to the final illustration (Phase 4: Meta), we use metaphor and narrative to depict the mediation role of the DbA in create process. Imagine a person standing before a wall. The DbA acts as a foundation of knowledge, elevating their perspective and enabling them to gain a bigger vision and inspiration. They then select content that aligns with their intrinsic value in order to generate ideas and begin prototyping. The DbA system then provides relevant information and details, assists with completing multiple specific operations, and subsequently transfers knowledge for usability evaluation.
    \textbf{See Table~\ref{tab:Literature_Distribution} for the classification of related papers corresponding to each stage.}}
    \Description{}
    \label{fig:Create Process}
\end{figure*}

\vfill
Drawing on prior work concerning the creative process~\cite{chung2021intersection, hsueh2024counts}, we analyze how DbA techniques operate across distinct stages of creation. Although many works span multiple stages, we organize our analysis into four phases comprising seven concrete stages, as shown in Fig.~\ref{fig:Create Process}.To clarify the literature distribution, we summarize the specific DbA techniques and related works for each stage in Table~\ref{tab:Literature_Distribution}. Within each stage, we describe cases according to the level of human-AI collaboration: \textit{Assist}, \textit{Augment}, and \textit{Automate}. \rv{As defined in Section 3, the systems examined in the current corpus range from Level of Automation (LoA) 2 to 7~\cite{parasuraman2000model}. Currently, no DbA systems achieve LoA 8--9, primarily because complex design tasks typically require human oversight and cannot yet be handled fully independently.}


\rv{To illustrate the mediating role of DbA throughout this workflow, we employ a narrative illustration (Fig.~\ref{fig:Create Process}).  In \textit{``Vision''}, DbA expands cognitive horizons by revealing hidden perspectives, as illustrated by individuals standing behind a wall and unable to see the distant landscape. In the \textit{``Inspiration''} phase, it enables access to diverse inspirational stimuli. During \textit{``Ideation''}, DbA facilitates the filtering and convergence of heterogeneous inputs into coherent concepts. In the \textit{``Prototype''} stage, the system recommends professional practices for optimizing methods. Subsequently, in the \textit{``Fabrication''} stage, DbA integrates with machines and systems to offer experiential assistance and knowledge transfer for specific implementations. Finally, the \textit{``Evaluation''} and \textit{``Meta''} stages both assist in achieving experiential migration.}


\begin{table}[!ht]
  \centering
  \caption{Classification of related papers across the four phases and seven stages shown in Fig.(\ref{fig:Create Process}).}
  \label{tab:Literature_Distribution}
  \footnotesize 
  \renewcommand{\arraystretch}{1.25} 
  
  \begin{tabular}{ll p{3.8cm}}
    \toprule
    \textbf{Phase} & \textbf{Stage} & \textbf{Related Papers} \\
    \midrule
    
    \multirow{2}{*}{Product Definition} 
      & Vision & \cite{zhang2017systematic, he2019mining, li2010agentsinternational, yu2014distributed, chan2014conceptual, christensen2016creative, gavetti2005strategy, moreno2014fundamental} \\
      & Inspiration & \cite{zhu2023design, emuna2024imitation, murphy2014function, moreno2014fundamental, chen2024asknaturenet, srinivasan2024improving, kang2025biospark, jiayang2023storyanalogy, warner2023interactive, bitton2023vasr} \\
    \midrule
    
    \multirow{2}{*}{Product Ideation} 
      & Ideation & \cite{kim2014bayesian, han2018computational, he2019mining, luo2021guiding, bettin2023pedagogical, Ju2025toward, kang2025biospark, zhai2020applying, coley2019robotic, zhang2020deep} \\
      & Prototype & \cite{masson2025textoshop, liu2023smfm, chen2021umitation, luo2019computer, fischer2024nerf, jin2022visual, gong2023toontalker} \\
    \midrule
    
    Product Implementation 
      & Fabrication & \cite{you2018design, jin2023deep, dimassi2023knowledge, emerson2024anther, schulz2014design, khosravani2022intelligent, zhai2020applying, an2020bim, gonzalez2018energy, zhang2020deep, lupiani2017monitoring} \\
    \midrule
    
    \multirow{2}{*}{Product Evaluation} 
      & Evaluation & \cite{dougan2022predicting, bhavya2023cam, boteanu2015solving, hill2019learning, kang2025biospark, emerson2024anther, malone2017putting, thomas2013extending, an2020bim, coley2019robotic, gonzalez2018energy} \\
      & Meta & \cite{andriani2025perfume, cao2025medai, thomas2013extending} \\
    \bottomrule
  \end{tabular}
\end{table}

\subsection{Phase 1: Product Definition}
\rv{In our framework, the \textit{Product Definition phase}, as illustrated in Fig.~\ref{fig:Create Process}, is divided into two stages. The \textbf{Vision} stage delivers directional inspiration to clarify vague problem directions, while the \textbf{Inspiration} stage generates heuristics and stimuli to enrich problem-framing.}

Specifically, the core mission of this phase is to precisely define problems by revealing structural contradictions rather than relying on superficial statements~\cite{simon2019sciences, hsueh2024counts}. Poorly defined problems risk misaligned solutions, wasted resources, and reliance on entrenched methods that restrict creative exploration~\cite{wills1994towards}. DbA techniques provide alternative perspectives for problem definition by transferring indirect experiences and knowledge from other domains, helping designers: (1) move from a constrained vision (``under the wall'') to a broader horizon (``beyond the wall'') in the \textit{Vision} stage; and (2) generate design heuristics and stimuli to expand problem framing in the \textit{Inspiration} stage. Integrating DbA into these two stages supports attribution, inference, and internal vision and mission building~\cite{hewstone1990ultimate}, especially when technological knowledge is fragmented and distributed across domains~\cite{li2010agentsinternational}. \rv{However, within this phase, DbA research has yet to fully explore how to translate tacit knowledge and experience into structured representations. This is largely because the knowledge required to define problems is often highly fragmented~\cite{rigaud2022exploring}.}

\textbf{\underline{Vision.}} DbA technology at this stage aims to provide directional inspiration and strategic support for addressing ill-defined problems, such as those in market research or risk prediction. At the \textit{\textbf{Assist}} level, DbA-based recommendation systems retrieve analogous product features to guide software development directions~\cite{zhang2017systematic}; similarly, DbA-enabled multimodal mining of case libraries supports the early design of transportation systems~\cite{he2019mining}.
At the \textit{\textbf{Augment}} level, multi-agent systems combine knowledge bases, simulations, and fuzzy logic to help forecast investment strategies~\cite{li2010agentsinternational}. Collaborative DbA systems also demonstrate how participants with diverse perspectives can co-construct multiple visions alongside AI agents~\cite{yu2014distributed, chan2014conceptual, christensen2016creative}.
At the \textit{\textbf{Automate}} level, scenarios apply DbA methods to generate heuristic directions autonomously; for instance, leveraging DbA-driven topic modeling in firm management assessment to inform policy strategy~\cite{gavetti2005strategy} and supporting problem definition for service design via DbA-based reasoning~\cite{moreno2014analogies}.

{
\sloppy
\textbf{\underline{Inspiration.}} This stage leverages DbA to provide inspirational stimuli and design heuristics. At the \textit{\textbf{Assist}} level, approaches include extracting bioTRIZ-inspired innovation stimuli through functional case modeling~\cite{zhu2023design}, developing semantic retrieval and reasoning models for mining biological analogies~\cite{emuna2024imitation}, employing hybrid vector models for functional search~\cite{murphy2014function}, and constructing rule-based case libraries for solution retrieval~\cite{moreno2014fundamental}. Additionally, tools such as \textit{AskNatureNet} facilitate divergent thinking by visualizing structured knowledge~\cite{chen2024asknaturenet}. At the \textit{\textbf{Augment}} level, \textit{AnalogiLead} enables linguistic chunking and recombination to foster new ideas, like ~\cite{srinivasan2024improving}, while \textit{BioSpark} integrates visual and textual stimuli to trigger inspiration~\cite{kang2025biospark}. At the \textit{\textbf{Automate}} level, DbA systems generate creative analogies with minimal human intervention. Examples include \textit{StoryAnalogy}, which utilizes large language models to mine and generate narrative-level analogies~\cite{jiayang2023storyanalogy}; \textit{VST}, which automates the flexible transfer of visual styles between vector graphics~\cite{warner2023interactive}; and \textit{VASR}, which synthesizes complex analogical images with contextual awareness~\cite{bitton2023vasr}.
}

\subsection{Phase 2: Product Ideation}

\rv{The \textit{Product Ideation} phase showcases DbA's capacity to translate diverse references into actionable concepts. Within this phase, the \textbf{Ideation} stage serves as a potent mechanism for integrating user intent, leveraging DbA to consolidate divergent inspirations into concrete, feasible design proposals. The \textbf{Prototype} stage further refines these proposals, elaborating on their details to expand possibilities for comprehensive and robust solutions.}

In detail, this phase centers on transforming problem definitions and exploratory analysis into feasible design solutions. As the ideation phase commits 70\% to 80\% of project costs, the quality of decision-making during this period critically determines the performance and sustainability of the product~\cite{saravi2008estimating}. \rv{Research indicates that DbA systems effectively reduce resource waste in design outputs and minimize design fixation~\cite{o2015toward}. However, regarding generative AI-driven systems, the consumption of computational power and physical materials remains an area requiring further exploration.} During ideation, by integrating domain knowledge and experience repositories~\cite{jiang2022data}, DbA supports the abstraction of source problems and the generation of target solutions, enabling goal-oriented innovations that address user needs~\cite{fu2015design}. In the subsequent \textit{Prototype} stage—the transition from ideation to implementation—DbA assists in structural refinement, aesthetic enhancement, and scheme polishing. However, current DbA technology focuses extensively on divergent thinking while often overlooking the implementation of convergent thinking, which holds potential for guiding design decisions and mediating user intention.

\textbf{\underline{Ideation.}} Unlike inspiration, ideation emphasizes design convergence, where DbA assists users in synthesizing multiple inspirations into concrete ideas by integrating user intentions. At the \textit{\textbf{Assist}} level, DbA supports designers through methods such as Bayesian case-based modeling for prototype generation~\cite{kim2014bayesian} and ontology-driven computational tools that identify structural similarities across domains~\cite{han2018computational}. Similarly, He et al. achieve directed design convergence through conceptual space mining~\cite{he2019mining}, while Luo et al. stimulate design ideation by regulating knowledge distance~\cite{luo2021guiding}. 
At the \textit{\textbf{Augment}} level, DbA enhances creativity with systems such as \textit{Pedagogical Prism} for domain isomorphic analogies~\cite{bettin2023pedagogical} and \textit{EmoSync}, which enables personalized empathy via affective reasoning~\cite{Ju2025toward}. Likewise, \textit{BioSpark} offers analogy suggestions that users can interactively adjust, gradually maturing early concepts into complete ideas~\cite{kang2025biospark}. 
At the \textit{\textbf{Automate}} level, DbA autonomously generates solutions; examples include case-based methods for mapping irrigation scheduling in viticulture~\cite{zhai2020applying}, robotic workflows for retrosynthesis prediction to automate organic compound synthesis~\cite{coley2019robotic}, and deep learning-based decision-making modules for digital twin manufacturing~\cite{zhang2020deep}.

\textbf{\underline{Prototype.}} In this stage, DbA assists users in refining, upgrading, and iterating predefined ideas. At the \textit{\textbf{Assist}} level, examples include \textit{Textoshop}, which adapts image editing logic to text editing~\cite{masson2025textoshop}, the use of semantic vectors for 4D printing material recommendation, and SMFM-based retrieval of product details for prototypes~\cite{liu2023smfm}.
At the \textit{\textbf{Augment}} level, DbA expands design possibilities through tools such as \textit{Umitation} for extracting UI behaviors from target websites for front-end development~\cite{chen2021umitation}, \textit{InnoGPS} for patent visualization and function detail recommendation~\cite{luo2019computer}, and \textit{Xcreation} for generating and polishing narrative illustrations.
At the \textit{\textbf{Automate}} level, systems employ automated material and modeling analogy transfer techniques, including NeRF-based 3D attribute transfer~\cite{fischer2024nerf}, CNN-HGCN pipelines for 2D visual analogy~\cite{jin2022visual}, and Transformer-based facial animation retargeting in \textit{Toontalker}~\cite{gong2023toontalker}.



\subsection{Phase 3: Product Implementation} 
\rv{In the \textit{Product Implementation} phase, the \textbf{Fabrication} stage highlights how DbA embeds knowledge-driven analogies into tangible production practices. By supporting the reuse of fabrication experience, standardizing niche manufacturing practices, facilitating process planning, and enabling case transfer, DbA empowers the final performance of manufacturing content, improves efficiency, and enhances creativity.}

Specifically, this phase focuses on transforming concrete ideas into tangible outputs through manufacturing, production, and deployment. While DbA research has been predominantly applied to early-stage creative ideation, relatively few works address implementation. This disparity reveals a critical epistemological gap: the challenge of translating highly discrete, case-specific operational knowledge into structured forms~\cite{rigaud2022exploring}. Without systematic documentation, invaluable experiential and embodied knowledge—ranging from replicable techniques~\cite{schulz2014design} and niche manufacturing practices~\cite{scopigno2017digital} to cultural craft expertise~\cite{vaisey2009motivation}—faces the imminent risk of erosion.  However, DbA holds significant promise in this domain: in intelligent manufacturing, DbA-based cross-domain mapping can balance fabrication creativity and accuracy by embedding structural transfer into automated production lines~\cite{schulz2014design}; in cultural heritage preservation, DbA can support the digitization and revitalization of craftsmanship by encoding tacit knowledge into standardized, computable forms for preservation and reuse~\cite{scopigno2017digital, jiang2022data}.

\textbf{\underline{Fabrication.}}
This stage spans manufacturing, development, production, and deployment. At the \textit{\textbf{Assist}} level, DbA facilitates the capture and reuse of prior operational knowledge. For instance, You et al. embed craft data into a geotechnical decision support system to guide detailed implementation choices~\cite{you2018design}. Similarly, Jin et al. leverage deep learning to assist buckling-guided assembly design for 3D frame structures~\cite{jin2023deep}, helping engineers reduce trial-and-error~\cite{dimassi2023knowledge}.
At the \textit{\textbf{Augment}} level, DbA enhances human expertise through interactive and data-driven transfer. For example, \textit{Anther} applies cross-domain video tutorial retrieval to transfer niche craftsmanship techniques~\cite{emerson2024anther}. Schulz et al. develop a system capable of designing, assembling, and manufacturing 3D products using example databases~\cite{schulz2014design}, while Khosravani et al. deploy knowledge-based DbA for injection molding, integrating domain knowledge into intelligent manufacturing lines~\cite{khosravani2022intelligent}.
At the \textit{\textbf{Automate}} level, DbA enables autonomous production workflows. The case-based reasoning system by Zhai et al. for agricultural irrigation can be embedded in smart city infrastructures~\cite{zhai2020applying}. An et al. develop a BIM-based automated manufacturability checking system for wood frames~\cite{an2020bim}, while other systems achieve intelligent management using technologies such as Case-Based Reasoning and deep learning, covering fields ranging from energy supply to elderly monitoring~\cite{gonzalez2018energy, zhang2020deep, lupiani2017monitoring}.



\subsection{Phase 4: Product Evaluation} 
\rv{The \textit{Product Evaluation} phase is divided into two stages: \textbf{Evaluation} and \textbf{Meta}. Together, they reveal how DbA not only supports the systematic assessment of individual products but also orchestrates design knowledge across domains and lifecycles. This composite role strengthens product sustainability throughout the entire lifecycle~\cite{wills1994towards}, enabling both localized optimization and global management of design processes.}

This phase addresses the monitoring and evaluation of products throughout their lifecycle, employing metrics such as user feedback, risk assessment, market response, projected outcomes, and policy compliance to enable optimization, prevention, explanation, alerting, and maintenance~\cite{moreno2014fundamental}. Yet, many products lack systematic evaluation during development, with assessments often delayed until post-deployment~\cite{zhong2017intelligent}. DbA techniques can mitigate this by enabling concurrent evaluation during development through data-driven case analysis and fuzzy logic assessment, thereby enhancing sustainability and long-term performance~\cite{liao2021priming}.


\textbf{\underline{Evaluation.}} DbA supports evaluation by predicting outcomes, providing critiques, and generating feedback on completed designs. At the \textit{\textbf{Assist}} level, DbA helps assess performance by drawing on prior cases or structured functions. For instance, Arditi et al. predict the outcomes of construction litigation using case data~\cite{dougan2022predicting}. Additionally, \textit{CAM} employs evaluation functions to assess the quality of creative analogy~\cite{bhavya2023cam}; Boteanu et al. interpret analogy meanings and questions through semantic networks~\cite{boteanu2015solving}; and Hill et al. test analogy precision by contrasting abstract relational structures~\cite{hill2019learning}.
At the \textit{\textbf{Augment}} level, DbA enhances human judgment through interactive evaluation. \textit{BioSpark} automatically generates solution recommendations and feedback for ideas~\cite{kang2025biospark}, while \textit{Anther} computes conceptual distance and ``concept gravity'' scores to evaluate analogical similarity across niche fabrication cases~\cite{emerson2024anther}. In business domains, Malone et al. apply contest network models to forecast profit and equity distribution in supply chains~\cite{malone2017putting}.
At the \textit{\textbf{Automate}} level, DbA systems conduct large-scale evaluation autonomously. For example, Thomas et al. extend and automate the optimization and generation of maintenance requirements for software and mechanical systems using System Theoretic Hazards Analysis (STHA)~\cite{thomas2013extending}. An et al. perform automated manufacturability checks~\cite{an2020bim}, with similar approaches applied in related studies~\cite{coley2019robotic, gonzalez2018energy}.


\textbf{\underline{Meta.}}
Beyond product assessment, DbA supports the global management of design processes by aligning workflows, knowledge systems, and interdisciplinary collaboration. At the \textit{\textbf{Assist}} level, DbA facilitates the transfer of end-to-end frameworks across industries. For example, Andriani et al. map perfume production workflows onto winemaking practices, demonstrating how tacit expertise can be reframed for distinct domains~\cite{andriani2025perfume}.
At the \textit{\textbf{Augment}} level, DbA enhances collaborative evaluation across disciplinary boundaries. Cao et al. develop an analogy tool that supports interdisciplinary collaboration between AI researchers and medical specialists by bridging terminology gaps to facilitate smooth communication~\cite{cao2025medai}.
At the \textit{\textbf{Automate}} level, DbA techniques automatically derive global design requirements from heterogeneous datasets. For instance, Thomas et al. identify iteration requirements for both software and mechanical development cycles~\cite{thomas2013extending}.



%% file: sections/6_Application.tex
\section{Applications}
\begin{figure*}[tp]
    \centering
    \includegraphics[width=0.85\linewidth]{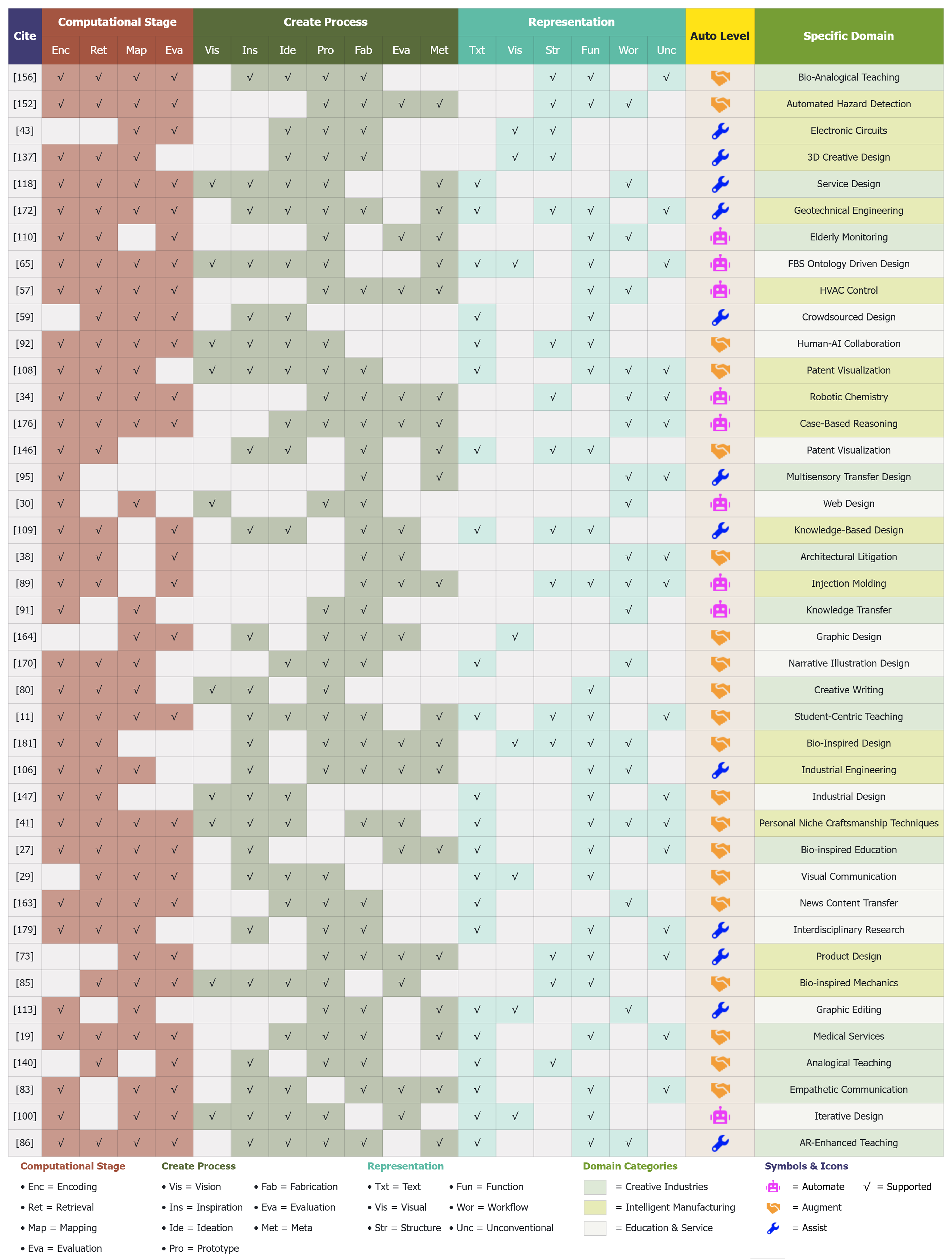}
    \caption{Part of systematic review result of Design-by-Analogy (DbA). The table classifies 41 surveyed systems and studies along five key dimensions: (1) the supported \textbf{\textit{Computational Stages}} stages~\cite{gentner2011computational, jiang2022data}, (2) the targeted \textbf{\textit{Create Process}} stages, (3) the \textit{\textbf{Representation}} types used, (4) the Level of \textbf{\textit{Automation}}, and (5) the \textit{\textbf{Specific Application}} Domain. The legend at the bottom provides a detailed explanation of all abbreviations, symbols, and color-coding.}
    \label{fig:application}
    \Description{}
\end{figure*}

This section examines the applications and challenges of DbA across three key sectors: \textbf{Creative Industries}, \textbf{Intelligent Manufacturing}, and \textbf{Education and Service Industries}. While DbA extends to domains such as pharmaceutical and algorithmic design, we focus on these three fields due to their extensive deployment and broad applicability. Fig.~\ref{fig:application} summarizes the 41 included systems, with specific applications detailed in the final column of the corpus overview. \rv{Following this review, we outline the specific ethical risks associated with implementing AI-driven DbA across these sectors.}

\subsection{Creative Industries} 
\begin{figure*}[t]
    \centering
    \includegraphics[width=1\linewidth]{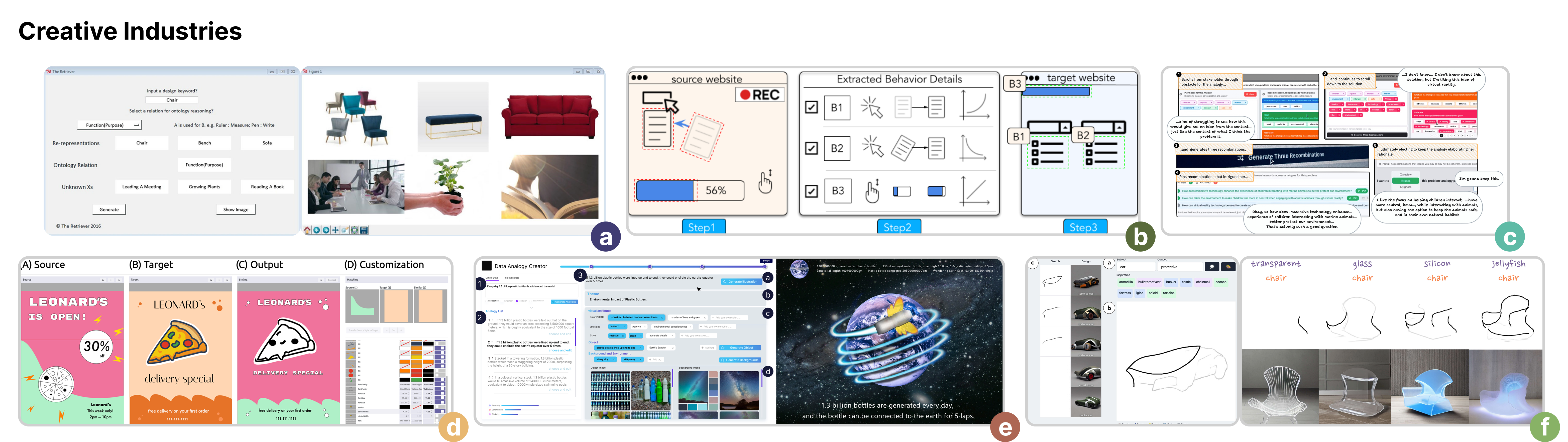}
    \caption{This figure showcases 6 cases deployed in Creative Industries in our corpus. (a) The retrieval tools for creative products based on analogical reasoning and ontology~\cite{han2018computational}. (b) Transfer source website UI action into target web development~\cite{chen2021umitation}. (c) Make creative analogies through chunking and recombination~\cite{srinivasan2024improving}. (d) Map source graphic style with target style~\cite{warner2023interactive}. (e) Leverage data analogy to enhance comprehension~\cite{chen2024beyond}. (f) Use analogical sketch support creative design~\cite{lin2025inkspire}.}
    \Description{}
    \label{fig:CreativeIndustries}
\end{figure*}

The UK Department for Culture, Media and Sport (DCMS) defines creative industries as activities rooted in individual creativity, skill, and talent, with the potential for wealth creation through intellectual property~\cite{miles2008hidden}. In our corpus, DbA applications in this sector encompass creative writing, graphic design, product design, 3D modeling, animation, web design, scientific researches, data visualization, and computer-aided design, whereas film and music remain largely absent. Representative cases are presented in Fig.~\ref{fig:CreativeIndustries}. Industrial outputs in this domain rely primarily on computational technologies and structured data, with limited influence from external non-creative factors such as political or force majeure events~\cite{miles2008hidden}. Our investigation identifies two core challenges for novices. First, high technical and aesthetic barriers in design software often prevent ideas from being fully realized or result in compromised outcomes~\cite{hsueh2024counts}. Second, emerging aesthetic and cultural trends are unevenly disseminated, as universal knowledge bases for design and creativity remain underdeveloped. DbA technologies address these challenges by providing scaffolding for creators at varying levels of expertise, automating stylistic and structural transfers, and sparking inspiration through analogical stimuli. For example, structured mapping computations enable the precise adaptation of graphical elements and styles, while access to external resources—such as patents, case studies, and near- or far-distance analogies—enriches idea exploration~\cite{luo2021guiding, song2020exploration, goucher2019crowdsourcing}. \rv{However, future DbA systems can only provide effective assistance when possessing a sufficiently deep understanding of user intent and the design problem itself. Furthermore, users must remain vigilant regarding the system's guidance mechanisms. Without this awareness, creators risk being misguided by persuasive generated content and imperceptible data biases, potentially abandoning critical thinking and stifling the creative aspects rooted in human intuition and artistic inspiration.} Ultimately, when designed responsibly, DbA tools aim not only to improve efficiency and quality but also to reinforce a sense of agency and ownership throughout the creative process.
 

\subsection{Intelligent Manufacturing} 

\begin{figure*}[t]
    \centering
    \includegraphics[width=1\linewidth]{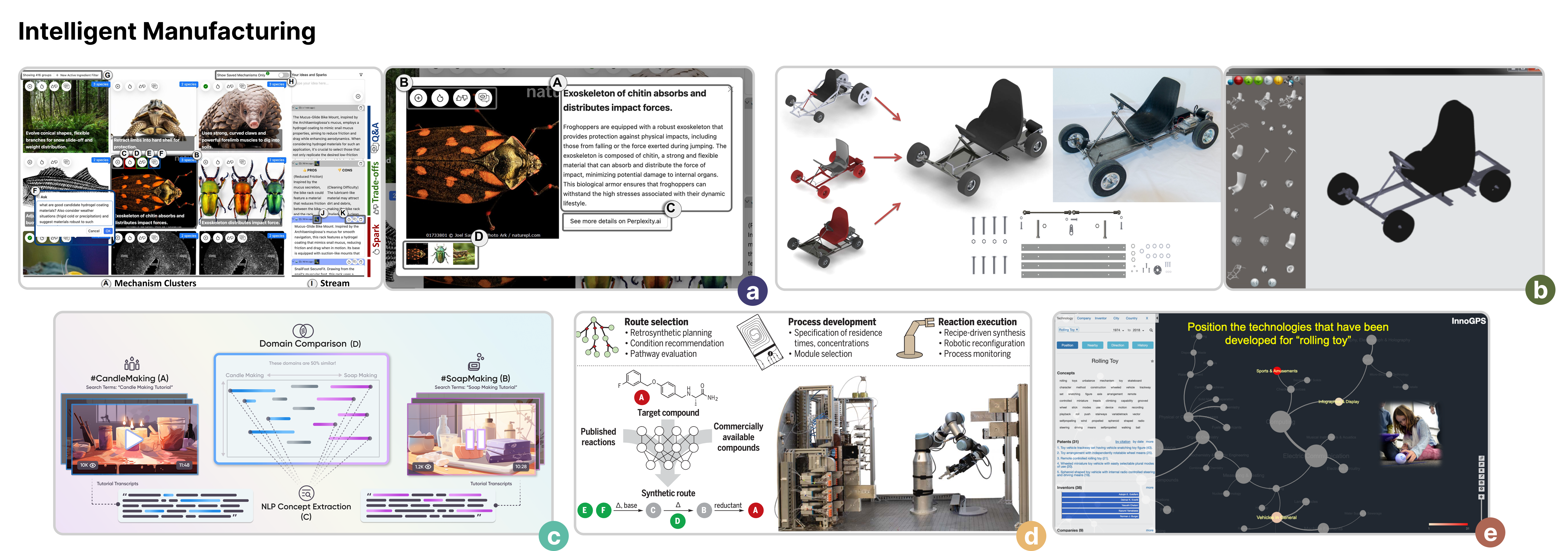}
    \caption{This figure illustrates 5 cases for intelligent manufacturing. (a) Augmented mechanical design creativity through bio-inspiration~\cite{kang2025biospark}. (b) Case-based design-to-manufacturing lifecycle system~\cite{schulz2014design}. (c) Process-oriented intelligent practice knowledge transfer~\cite{emerson2024anther}. (d) Chemical synthesis based on previous process transfer~\cite{coley2019robotic}. (e) GPS-like tool for manufacturing method exploration~\cite{luo2019computer}.}
    \Description{}
\label{fig:IntelligentManufacture}
\end{figure*}

In Industry 4.0, intelligent manufacturing refers to the integration of artificial intelligence (AI), the Internet of Things (IoT), and cyber-physical systems to transform traditional manufacturing resources into intelligent objects with perception, decision-making, and collaboration capabilities. This enables flexible data-driven production for mass customization~\cite{zhong2017intelligent, li2017applications}. In our corpus, DbA applications in this domain include biology knowledge-based design inspiration~\cite{kang2025biospark}, case-based design-to-manufacturing lifecycle systems~\cite{thomas2013extending, schulz2014design}, case-based cyber-physical systems~\cite{gonzalez2018energy}, and process-oriented intelligent knowledge transfer~\cite{emerson2024anther}. These applications span manufacturing method design~\cite{luo2019computer}, product innovation~\cite{liu2023smfm}, smart sensor and wearable device development~\cite{hong2024fishbone}, process planning~\cite{zhang2020deep}, additive manufacturing~\cite{hagedorn2018knowledge}, plastics processing~\cite{khosravani2022intelligent}, chemical or pharmaceutical synthesis~\cite{coley2019robotic}, energy-saving control systems~\cite{gonzalez2018energy}, and agriculture~\cite{zhai2020applying}, as shown in Fig.~\ref{fig:IntelligentManufacture}. DbA contributes to this field in three key ways. First, far-distance analogy transfers enable engineers to incorporate cross-domain knowledge into manufacturing processes, fostering innovative applications of manufacturing resources. Second, near-distance analogy transfers support foresight, sustainability, and efficiency by drawing on process planning experience and knowledge data, advancing the lifecycle servitization of manufacturing resources. Third, DbA provides a replicable pathway for codifying niche techniques, promoting innovation within communities of practice and enabling the archival of ambiguous or tacit knowledge~\cite{zhong2017intelligent}. Overall, integrating DbA in intelligent manufacturing offers novices heuristic resources and assists experts in enhancing complex problem-solving through function and structure migration. It optimizes the implementation of modular production units, flexible logistics planning, and mass personalized customization~\cite{li2017applications}, while improving energy and resource allocation to advance sustainable development. \rv{However, intelligent manufacturing continues to necessitate decision-making processes that integrate embodied experience~\cite{varela2017embodied} and tactile perception. To mitigate the potential erosion of human embodied cognition, it is crucial that manufacturing systems incorporate authentic human sensory inputs and embodied reasoning methods more profoundly.}

\subsection{Education and Service Industries}

\begin{figure*}[t]
    \centering
    \includegraphics[width=1\linewidth]{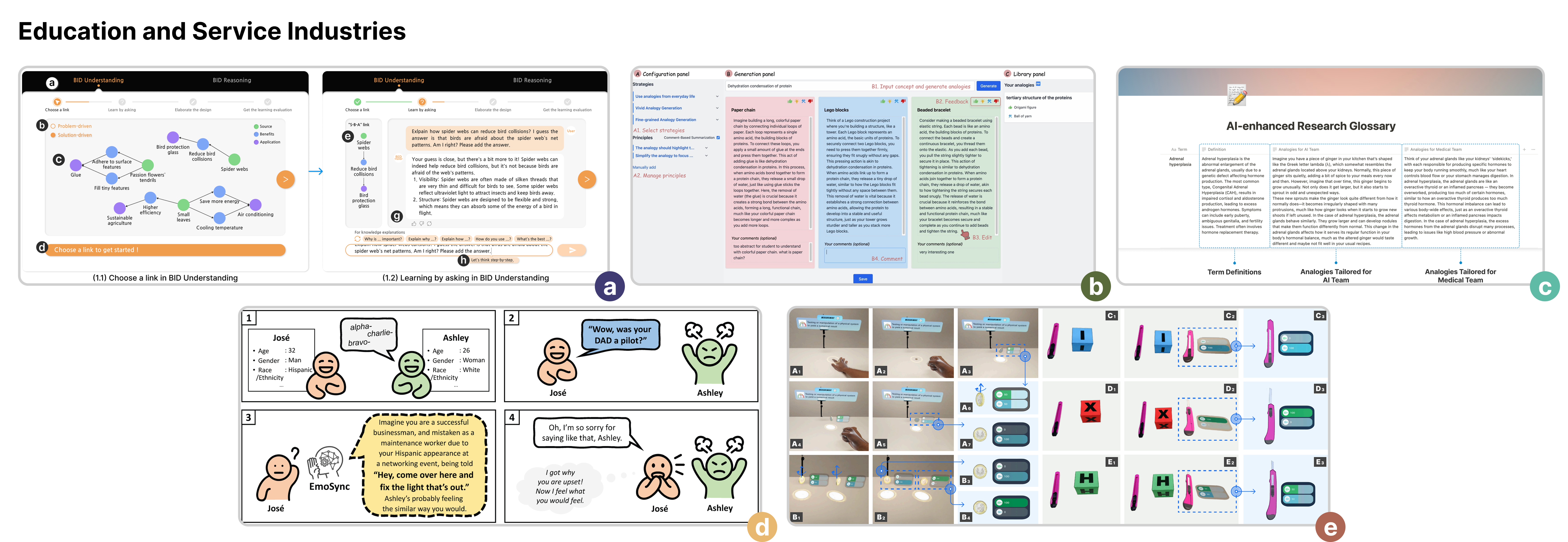}
    \caption{This figure illustrates 5 cases for Education and Service Industries. (a) Leverage the FBS structure of bio-inspiration data for BID education \cite{chen2024BIDTrain}. (b) Connect new concepts to familiar ideas in scientific teaching \cite{shao2025unlock}. (c) Construct cross-domain analogical terminology transfer logic for Med-AI collaboration \cite{cao2025medai}. (d) Generate personal analogy for affective empathy \cite{Ju2025toward}. (e) Explain quantum computing concepts through AR analogy \cite{karunathilaka2025intuit}. }
    \Description{}
    \label{fig:Education}
\end{figure*}

Analogical reasoning, the core principle of Design-by-Analogy (DbA), is also fundamental to learning and cognition~\cite{winston1980learning}, with deep roots in philosophy and extensive applications in pedagogy and mathematics~\cite{bettin2023pedagogical}. Unlike pure cognitive exercises, DbA applications examined in this educational context provide structured scaffolding for analogical application in real learning environments~\cite{winston1980learning, vattam2011dane, bettin2023pedagogical}. By systematically retrieving knowledge, experience, and data, and subsequently transferring them via domain-specific mapping rules to target scenarios, DbA tools empower educators to rapidly leverage adapted resources. This process advances students' conceptual understanding, metacognition, and interdisciplinary exchange~\cite{karunathilaka2025intuit, chen2024BIDTrain, ball2019advancing}. For example, \textit{\textbf{Pedagogical Prisms}} creates analogies based on students' unique backgrounds, experiences, and cultures to facilitate teaching~\cite{bettin2023pedagogical}. Additional representative cases are presented in Fig.~\ref{fig:Education}.

In the service industry, DbA plays an equally crucial role, facilitating transactional innovation across diverse cultures, domains, services, and communities through communication, collaboration, and co-creation. The core mechanism involves the explicit mapping of tacit knowledge, such as experiences and processes, enabling practitioners to integrate existing expertise with novel insights to address key challenges~\cite{lee2020customized, moreno2014analogies}. Its value is demonstrated in several ways. First, despite the highly context-dependent nature of service design, common methodologies like \textit{``touchpoint analysis''} and \textit{``service blueprints''} can be distilled to build experience repositories that support cross-scenario inspiration and solution recommendations~\cite{stickdorn2012service}. Second, transferring proven solutions from other domains can significantly enhance the practical efficacy of new service designs. Third, in a field that places a high premium on evaluation, DbA-based assessment methods hold significant potential for evaluating and predicting service efficacy. \rv{However, complex educational and service scenarios encompass personalized user experiences, consequential life decisions, idiosyncratic cases, and unique educational objectives. When imparting knowledge and delivering services, rigorous attention to specific contexts and the regulation of data bias~\cite{Ju2025toward} must be integral to the process. This approach helps mitigate the risk of oversimplifying the complex challenges individuals face—a pitfall often resulting from the application of one-size-fits-all templates.}

%% file: sections/7_Findings.tex
\section{Findings}
This section synthesizes insights derived from the preceding analysis, addressing four pivotal findings.

\subsection{\rv{Understanding DbA as Technological Mediation in Human-AI Collaboration (RQ1)}}
Drawing from design philosophy, Verbeek's theory of technological mediation frames technology as an active mediator of human-world interactions, embedding moral dimensions within its functionality~\cite{verbeek2006materializing}. In this view, technology co-constitutes human experience, influencing not only ``what we do'' but also ``how we perceive value''~\cite{verbeek2011moralizing}.


We argue that DbA exemplifies this mediation role. Its mechanisms of knowledge mapping and contextual transfer actively shape design cognition. First, DbA guides reasoning while preserving human agency: designers' intrinsic knowledge and values steer the analogical trajectory, positioning them as the primary authors of outcomes~\cite{chan2011benefits}. Second, AI-driven DbA can function as an ethical mediator by curating analogies or filtering harmful content, thereby embedding ethical boundaries directly into the creative process~\cite{liao2021priming, smits2019values}.

By disrupting cognitive rigidity, DbA activates core human values, prompting designers to actively construct meaning, exercise judgment, and engage curiosity—rather than passively executing steps. This active engagement reinforces their subjectivity within the creative process~\cite{schecter2025role, smits2019values, shao2025future}. 
For instance, when designers actively draw analogies through DbA methods between ecosystems and urban planning, they foster holistic, human-centered solutions that transcend mere passive replication. Such processes cultivate a sense of ownership, mastery, and purpose—intrinsic motivations essential for ethical and innovative practices.
However, our review also reveals a significant gap between this conceptual potential and real-world adoption. Many engineering communities remain oriented toward automation-centric paradigms~\cite{smits2019values, schecter2025role, wills1994towards}, where creativity is undervalued and technology is framed primarily as a labor-saving mechanism. This epistemological stance obscures the distinction between DbA and retrieval-based approaches such as CBR, leading to resistance toward systems that foreground human judgment and value mediation~\cite{wills1994towards, smits2019values}. Bridging this gap requires reframing DbA not as a substitute for human creativity, but as a value-laden partner that reallocates cognitive effort toward higher-order reasoning.

Critically, our findings also highlight that DbA’s mediating power is ethically ambivalent. While the democratization and transfer of analogical knowledge can empower designers and elevate creative agency, the same mechanisms introduce dual-use risks. The abstraction and portability that make DbA effective for innovation may also enable the misuse of cross-domain knowledge in harmful contexts, including the design of dangerous artifacts or processes. DbA does not inherently promote beneficial outcomes; without deliberate constraints, it may equally facilitate reckless or malicious applications. This underscores the necessity of embedding value-sensitive design principles, domain governance, and rigorous pre-deployment assessments to ensure that ``liberation'' from routine work does not translate into disempowerment or harm~\cite{de2018can, smits2019values}.

Taken together, the literature reviewed here positions DbA as more than a technical method: it is a form of technological mediation that reshapes agency, responsibility, and value expression in Human–AI collaboration. Realizing its promise depends not only on algorithmic capability, but on conscious design choices that align analogical power with human dignity, ethical accountability, and long-term societal impact.

\subsection{\rv{The Rationale for Broadening the Application Scope of DbA (RQ2)}} 

DbA extends beyond its traditional design applications because it addresses the limitations of other computational reasoning approaches. Case-Based Reasoning (CBR), while efficient and widely adopted, largely overlooks human agency and cultural context~\cite{wills1994towards}. This rigidity makes CBR effective for routine tasks but less suitable for value-driven or emerging domains~\cite{goel2012cognitive}. In contrast, DbA retains the structured mapping logic of CBR while introducing critical flexibility. It allows users to select and refine analogical mappings, integrates cultural and contextual signals, and adapts to different stages of the creative process. These features turn user choices—such as selecting analogies—into active contributions that shape design outcomes, often yielding more diverse and personalized results~\cite{chan2011benefits, moreno2015step, moreno2016overcoming}.

Furthermore, DbA challenges the prevailing efficiency-centricism of the industrial sector. By virtue of its inherent structured mapping logic, DbA seamlessly accommodates existing industrial process planning and manufacturing method transfers while imbuing industrial operations with an indispensable creative foundation. As industrial values transition from a strict focus on efficiency toward personalization, the strategic integration of DbA holds the potential to realize the visions of mass customization and Manufacturing-as-a-Service (MaaS)~\cite{zhong2017intelligent}. In the AI era, DbA also leverages the emergent capabilities of large models: these models can act as backend engines for combinational creativity~\cite{peng2025probing}, retrieving expert knowledge while embedding cultural signals derived from online data. This positions DbA as a mediating technology that connects structured reasoning with human values and evolving cultural practices.

However, while the vision of this methodology has been articulated, its deep integration into industrial systems in the future will inevitably introduce exponentially increasing noise and risks into currently stable and harmonious industrial processes. For instance, when employing generative intelligence for parameter migration in large CNC machine tools, inaccuracies in generated content could result in material damage at best and pose serious safety hazards at worst. Simultaneously, new phenomena such as human arrogance stemming from augmentation and intellectual hubris may foster novel social ideologies. Deploying this method in the real world carries heightened risks due to the diversity and fluidity of its processes compared to traditional approaches, thereby elevating the comprehensive competency requirements for human-machine interaction designers. It is foreseeable that the impact of such systems on the traditional industrial order will be significant. Industries must seek more stable and secure breakthrough solutions while respecting and supporting human creativity.


\subsection{\rv{Common Guidance for DbA Techniques (RQ3)}}
Despite the diversity of implementations, common principles for DbA can be distilled along four dimensions: \textbf{user}, \textbf{data}, \textbf{algorithms}, and \textbf{interaction}.

\textbf{\underline{User.}} DbA systems should prioritize recognizing human intention rather than assuming complete automation. 

\begin{itemize} [leftmargin=1em, labelsep=0.5em, nosep]
\sloppy
\setlength{\emergencystretch}{3em}
\item \textbf{Recognize User Intention through Cognitive Methods.} \rv{For instance, drawing upon cognitive models such as the Technology Acceptance Model~\cite{davis1989perceived}, systems can infer user intentions and attitudes by analyzing actual behaviors—specifically, clickstream content and navigation trajectories.} Furthermore, DbA should incorporate insights from neural and cognitive science to detect shifts in users' creative states~\cite{goel2012cognitive}. For example, during formative research, EEG, eye tracking, or behavior logs (e.g., long pauses, repeated clicks, task switching) can be linked to specific cognitive states~\cite{borgianni2020forms, goucher2019neuroimaging}. Such mappings allow the system to infer when support is needed and adapt assistance accordingly, enabling sustained creative work.
\item \textbf{Implement Differentiated Scaffolding for Experts and Novices.} Empirical evidence confirms that experts and novices employ distinct neural pathways and cognitive structures during problem-solving~\cite{chai2015behavioral, goucher2019neuroimaging}. Future creativity support systems should avoid a ``one-size-fits-all'' approach. Novices may benefit from near-distance, structured resources (e.g., based on Function-Behavior-Structure models), while experts may require far-distance, abstract stimuli in the early stages, supplemented with metacognitive tools to activate their intrinsic knowledge~\cite{ball2019advancing, viswanathan2016study}. During refinement, the system could then deliver highly specific, near-distance stimuli to facilitate detailed execution. 
\end{itemize}

\textbf{\underline{Data.}} We propose three guidelines for DbA-related data design and processing. 
\begin{itemize} [leftmargin=1em, labelsep=0.5em, nosep]
\sloppy
\item \textbf{Encode Experience into Structured Data.} Structured data should extend beyond parameters to include experience, trajectory, causality, and phenomena~\cite{bandemer2012fuzzy}. For example, frameworks such as Function-Behavior-Structure~\cite{chen2024toward}, Input-Processing-Output~\cite{boell2012conceptualizing}, and Vision-Language-Action~\cite{zitkovich2023rt} can serve as ontological templates for building reusable knowledge bases. 
\item \textbf{Enhance Data Reusability and Cross-Scenario Adaptability.} Instead of fragmented databases, DbA systems should organize information into domain-specific modules (e.g., a ``manufacturing process data module'' containing equipment parameters, operational workflows, and fault resolution protocols). These modules can be embedded across systems using standardized interfaces and mapping rules, enabling reuse in contexts like MaaS (Manufacturing as a Service) or model fine-tuning~\cite{zhong2017intelligent}. 
\item \textbf{Use Synthetic Data to Enrich Analogy Databases.} When high-quality data are scarce, synthetic generation can augment coverage and rebalance distributions~\cite{paulin2023review}. For example, rare techniques in niche manufacturing can be supplemented through rule-based or model-driven generation, expanding the analogy search space without incurring the costs of extensive data collection.
\end{itemize}


\textbf{\underline{Algorithm.}} We identify four key guidelines for DbA algorithm design. 
\begin{itemize} [leftmargin=1em, labelsep=0.5em, nosep]
\sloppy
\item \textbf{Adopt Context-Adaptive Selection and Fusion.} A context-adaptive strategy for algorithm selection is crucial, as different methods and models present distinct trade-offs. 
For example, \textit{WordTree} effectively breaks design fixation but imposes a high cognitive load, while \textit{SCAMPER} offers systematic guidance with a lower risk of fixation~\cite{moreno2016overcoming}. Similarly, \textit{symbolic} models offer speed for well-defined tasks; \textit{connectionist} models excel with structured data but struggle with complexity or data scarcity; and \textit{hybrid} models, while robust, require complex engineering architectures. Therefore, systems should help diagnose context and strategically combine methods~\cite{jiang2022data}.

\item \textbf{Balance Far- and Near-Distance Analogies.} Far-distance analogies broaden the scope during early vision and inspiration stages~\cite{kang2025biospark}, while near-distance analogies support refinement in ideation, prototyping, and fabrication~\cite{tseng2008role, dimassi2023knowledge}. The evaluation phase often involves a collaborative mix of both~\cite{thomas2013extending}. Beyond design stages, the optimal distance is also modulated by specific goals (quantity favoring near-distance, quality favoring far-distance), user expertise (novices requiring near-field, experts far-field), and personality traits, which require tailored approaches~\cite{chan2011benefits, song2018characterizing, chai2015behavioral}.

\item \textbf{Enable Multimodal and Dynamic Representation Mapping.} A core feature of DbA is that representation mapping is not a fixed, automated process but a dynamic, human-integrated one. Mapping should flexibly integrate semantics, function, structure, and appearance using multimodal similarity measures~\cite{hill2019learning, boteanu2015solving}. It should also adapt dynamically to user identity, creative stage, and evolving intent, embedding human perception and experience into the algorithm~\cite{cao2025medai, dimassi2023knowledge}. 

\item \textbf{Support Expert-Led Mapping Rules.} 
Domain experts should author and refine mapping rules to capture tacit knowledge and domain-specific creativity. This is especially critical in large-model-based systems where content and interaction are tightly coupled, requiring a holistic design that bridges product and algorithm development. 
\end{itemize}


\textbf{\underline{Interaction.}} We outline three principles for DbA interaction design.
\begin{itemize} [leftmargin=1em, labelsep=0.5em, nosep]
\sloppy
\item \textbf{Use Interfaces as Metacognitive Scaffolds.} Rather than traditional ``toolbox'' manipulation of graphical primitives (e.g., points, lines, planes), interfaces should externalize structural logic through visual metaphors such as graphs~\cite{yan2023xcreation, huang2024field}, tree diagrams~\cite{chen2024BIDTrain, chen2024asknaturenet}, and interactive canvases~\cite{lin2025inkspire, masson2025textoshop}. This makes abstract mappings explicit and navigable.
\item \textbf{Design for Process Awareness and Fluidity.} The interface should dynamically adapt to the creative workflow, providing seamless transitions between stages. This implies supporting extensive information intake during early exploration, enabling meticulous refinement in mid-stage ideation, and offering dashboard-driven visualizations for late-stage evaluation~\cite{kang2025biospark, srinivasan2024improving}.
\item \textbf{Provide Subtle yet Effective Guidance.} Instead of prescriptive instructions, systems should leverage implicit interactions—such as iconography for analogy distance indicators, translucent overlays, and ephemeral cues—to nudge users toward promising analogical directions~\cite{ju2008design}. The key is to mitigate the risks of design fixation and outcome standardization often exacerbated by prescriptive guidance.
\end{itemize}



\subsection{\rv{Future Prospects for DbA Application and Research (RQ3)}}
In the age of generative AI, DbA's promise lies not in a monolithic application~\cite{hsueh2024counts} but in its role as a versatile intelligence layer~\cite{shao2025future}. Tailored to specific needs, DbA can embed innovative capabilities across diverse technological ecosystems~\cite{rigaud2022exploring, zhang2020deep}. 
Its potential applications span multiple scales: lightweight, cloud-based applications, such as interactive dashboards, employ analogy for strategic planning and market analysis; platform-level creative suites, including next-generation CAD software, integrate DbA to support analogical design~\cite{goel2012cognitive}; embedded models in robotics and manufacturing translate abstract parameters into operational commands for Manufacturing-as-a-Service (MaaS); and ambient or context-aware systems—ranging from projectors mapping analogies onto physical spaces to agentic schedulers orchestrating resources—bring analogy into the physical environment~\cite{wu2025agentic}. Collectively, these directions signal DbA's evolution from a standalone technique or tool to an underlying engine for innovation.


Despite its growing applications, several research gaps remain. 
First, there is a lack of structured public knowledge repositories, particularly for emerging domains like sustainability and the circular economy~\cite{liao2021priming}, limiting DbA in complex settings that demand heterogeneous data integration~\cite{moreno2014analogies, ross2022exploring}. Second, DbA's use in intelligent manufacturing is overly focused on conceptual design ~\cite{hsueh2024counts}, while its potential for the fabrication stage remains underexplored due to the difficulty of capturing tacit shop-floor knowledge ~\cite{emerson2024anther, ross2022exploring}. Third, interaction modalities are overwhelmingly confined to conventional 2D software, leaving the potential of immersive, tangible, or brain-computer interfaces as a vast, unexplored frontier. Finally, and most critically, there is a significant lack of rigorous empirical evaluation of AI-driven DbA outputs ~\cite{verhaegen2013refinements}. Without systematic studies of novelty, feasibility, and utility, claims of DbA's practical value cannot be substantiated ~\cite{liao2021priming}.

%% file: sections/8_Discussion.tex
\section{Discussion and Future Work}
Building on the preceding research, we integrate design intent, user perception, taxonomy, and ethical risks to provide an in-depth discussion on key topics, aiming to support responsible and human-centered innovation in future DbA systems.

\rv{\textbf{Ethical and Complexity Issues. }}
\rv{While DbA systems aim to preserve design intent, their future development must address three key ethical risks:
First, erosion of essential creative abilities\cite{kosmyna2025your}. Uncritical adoption fosters user dependency on automated tools, which can weaken intuitive and embodiment thinking skills. To counter this, systems should incorporate implicit metrics like cognitive activation \cite{greenwald1996three} and use interactive algorithms that encourage divergent thinking over strong tendencies.
Second, oversimplification of real-world complexity. Lacking nuanced context, DbA systems may cause designers to reduce complex problems to fit standardized inputs, ignoring real-world intricacies. This necessitates embedding insights from the empirical experience\cite{olschewski2024future}, design research, sociology\cite{chesluk2024systems} and other perspectives into the systems.
Third, deep embedding of data bias: DbA's high dependency on training data risks transferring and deeply embedding societal and cultural biases into designs. Solutions require interdisciplinary collaboration for data integrity, advanced bias detection and filtering technologies, and a critical examination of the complex interplay between values, knowledge, and bias.}

\textbf{The Complementary Nature of Human and AI Analogical Reasoning. }
While both AI and humans engage in analogical reasoning, they do so through fundamentally different principles. AI excels at identifying statistical correspondences across vast datasets, essentially performing complex pattern matching yet lacking semantic grounding and metacognitive awareness. In contrast, human analogy is a goal-directed act of sense-making rooted in contextual understanding~\cite{gentner1983structure, gentner2011computational, webb2023emergent}. This distinction highlights a cognitive gap: AI reasoning is performative, whereas human reasoning is intentional.


However, this distinction does not imply a simple ``creative'' versus ``non-creative'' binary. AI can generate novel associations that humans might overlook due to cognitive load or fixation~\cite{richland2013reducing, lu2023differences}, while humans contribute semantic validation, ethical judgment, and purposeful direction. The productive future for DbA lies in viewing them as complementary partners: AI acts as an expansive generator for candidate mappings, and humans serve as curators who filter, contextualize, and integrate them into innovation processes. Future work should focus on hybrid frameworks that couple algorithmic exploration with human interpretive depth.


\rv{This complementary perspective reinforces our definition of DbA as a technological mediation. Far from replacing human effort and opposing towards AI, DbA orchestrates a synergy between machine-scale retrieval and human meaning-making. By delegating structural search to AI while retaining semantic judgment for designers, DbA fulfills its core intent: bridging computational efficiency with the irreducible ``goal-directed'' nature of human creativity.}

\textbf{Knowledge Accumulation and Collaboration Across Time and Space.} 
\sloppy
\rv{The evolution of multi-modal technologies directly expands the scope of the Representation Taxonomy established in Section 4. While current DbA systems predominantly rely on \textit{Semantics \& Text} or \textit{Visual \& Appearance}, emerging tools promise to unlock richer, implicit dimensions. By capturing and preserving these ephemeral forms of knowledge, future DbA systems can evolve from static retrieval to the dynamic synthesis of tacit expertise and collective intelligence.}

Emerging technologies such as wearable devices and brain-computer interfaces (BCIs) connect human cognition directly with digital environments, transforming knowledge work from deliberate, manual retrieval into context-aware, dynamic exchanges~\cite{jensen2011using, moore2010applications}. For instance, BCIs could decode neural patterns to retrieve latent experiences, while wearables could interpret creative gestures to trigger cross-domain knowledge sharing in real time~\cite{birbaumer2006physiological, gao2021interface, nicolas2012brain}. These mechanisms dynamically reassemble fragmented knowledge from disparate times and places into a living ecosystem for computing, sharing, and accumulating individual and collective experience.

However, this vision raises challenges regarding privacy, ownership, and governance. Knowledge exchange cannot flourish if contributors fear the loss of control over sensitive data~\cite{jain2016big}. A robust solution requires a multifaceted strategy, beginning with technical safeguards like federated learning and differential privacy to enable knowledge transfer without exposing raw data~\cite{wu2012effect, grossman2004international}. This must be complemented by institutional frameworks, such as adaptive intellectual property rights and ethical guidelines, to clarify responsibilities in cross-domain exchange. Finally, community co-creation is essential, fostering open-source contributions of sanitized or high-quality synthetic datasets to enrich shared resources while mitigating risks~\cite{tamburri2019discovering, bauer2024comprehensive}.

A long-term research agenda involves engineering secure systems where tacit knowledge and creative trajectories are both protected and transmitted. Such systems would allow geographically and temporally distributed teams to access cross-cultural insights, enabling future generations to build upon the creative pathways of their predecessors.




\textbf{Designing for Human Agency. }
The integration of AI into creative workflows should cultivate, not diminish, human agency~\cite{hsueh2024counts, khanolkar2023mapping}.  
Effective systems balance two pillars: granting users meaningful control over operations and outputs, and fostering trust in the system's capabilities and boundaries. Ideally, DbA acts as transparent scaffolding, offloading cognitive overhead while reserving space for human intuition, critical judgment, and creative curation~\cite{rehm2017designing, frich2019mapping, chung2021intersection}. \rv{Realizing this vision requires the stage-specific support detailed in Section 5. Moreover, it necessitates deeper inquiry into human creative behavior and the adoption of extended design processes.}


This approach has profound implications for democratizing creativity by reallocating cognitive resources. By embedding expert knowledge into accessible tools, users are empowered to become ``full-stack designers''~\cite{bogenhold2014entrepreneurship}. For instance, a designer can apply analogical rules between biological forms and augmented manufacturing systems~\cite{yang2018recent}, bypassing tedious modeling to focus on higher-level tasks such as aesthetics, ethical impact, and value propositions~\cite{schecter2025role}. Here, external knowledge serves as inspiration rather than a directive, amplifying, rather than suppressing, the user's internal experience and intuition. 


However, such empowerment must be coupled with responsibility~\cite{smits2019values, rigaud2022exploring}. Systems should embed prudent evaluation mechanisms as guardrails. For example, a system could warn of potential risks by referencing analogous historical cases~\cite{thomas2013extending} or provide feasibility recommendations based on cross-domain performance data~\cite{andriani2025perfume}. 
Future research should examine how to quantify agency within DbA workflows, design adaptive feedback mechanisms that balance freedom and constraint, and evaluate long-term impacts on creativity and decision quality.

\balance

%% file: sections/9_Conclusion.tex
\section{Conclusion}
This systematic review examines the Design-by-Analogy (DbA) methodology driven by AI, underscoring its role as a technological mediator and a universal cognitive foundation for activating innate human capabilities and bridging cross-domain research. \rv{By situating DbA within the historical evolution of creativity support tools, this study elucidates its intervention in the creative process, specifically to mitigate design fixation potentially induced by oversimplified AI creative tools.} \rv{To address current research gaps, a thematic analysis was conducted on 1,615 publications, followed by an in-depth review of 85 core documents, acknowledging limitations in data collection and screening.} The results indicate that DbA supports four stages of the creative process through six categories of representations and technologies, with deployable applications across three distinct domains.
Specifically, we demonstrate DbA's potential for cross-modal knowledge transfer, computational integration, fuzzy data acquisition paradigms, and novel interaction modalities. \rv{Furthermore, we discuss the threefold ethical risks and their specific implications across various applications. Looking forward, DbA systems are poised to serve as technological mediators that reinforce cognitive agency centered on human values, while assisting humans in the critical application of AI analogical reasoning capabilities. This work provides a theoretical basis for fostering human a more mature confidence of creation.} Ultimately, these contributions support the next generation of responsible, human-centered design innovation.

%% file: sections/10_Acknowledgment.tex
\begin{acks}
    The authors would like to thank
the anonymous reviewers for providing constructive feedback. This work was supported by NSFC 62372327, NSF Shanghai 23ZR1464700. This work was also supported by the Fundamental Research Funds for the Central Universities and Tongji University Xinrui Young Talents Grant for Social Sciences and Humanities. 
\end{acks}